\documentclass[preprint,12pt,authoryear]{article}

\usepackage{amssymb}
\usepackage{amsmath}
\usepackage{tikz}
\usetikzlibrary{calc}

\usetikzlibrary{positioning}
\usetikzlibrary{arrows}
\usepackage{comment}
\usepackage{amsmath}
\usepackage{amssymb}
\usepackage{subcaption}
\usepackage{mleftright}
\usepackage{hyperref}
\usepackage{bm}
\usepackage{bbm}
\usepackage{setspace}

\usepackage{algorithm}
\usepackage{algorithmic}
\usepackage[autostyle]{csquotes}
\usepackage[title]{appendix}
\usepackage{pifont}
\usepackage{mathtools}
\usepackage{array}
\usepackage{natbib}

\usepackage{booktabs}
\usepackage{accents}

\newcommand{\mc}[1]{\mathcal{#1}}
\newcommand{\bs}[1]{\boldsymbol{#1}}

\newcolumntype{P}[1]{>{\centering\arraybackslash}p{#1}}
\usepackage{booktabs}
\usepackage{accents}
\usepackage{authblk}

\usepackage{xr}
\usepackage{cleveref}

\begin{document}

\title{Dynamic factor and double PCA models for partially observed survival curves: Forecasting demand in short-term rental markets} %% Article title
\author[1,2,*]{Marthe Elisabeth Aastveit}
\author[1]{Alex Lenkoski} 
\author[2]{Thordis Thorarinsdottir}

\affil[1]{Norwegian Computing Center, Oslo, Norway}
\affil[2]{Department of Mathematics, University of Oslo, Oslo, Norway}
\affil[*]{Corresponding author: Marthe Elisabeth Aastveit: aastveit@nr.no}
\date{}
%% Author affiliation
%\affiliation[label1]{organization={Norwegian Computing Center},%Department and Organization
%            addressline={P.O.Box 114 Blindern}, 
%            city={Oslo},
%            postcode={0314}, 
%            country={Norway}}
%\affiliation[label2]{organization = {Department of Mathematics, University of Oslo},
%  addressline={P.O.Box 1053 Blindern}, 
%            city={Oslo},
%            postcode={0316}, 
%            country={Norway}}            
%\cortext[cor1]{Corresponing author}

\maketitle
%% Abstract
\begin{abstract}
%% Text of abstract
 This paper develops prediction models for population-level survival curves observed over time and sampled from a heterogeneous mix of populations. We consider a discrete-time setting where each curve is only partially observed and forecasts of the remaining trajectory are needed for downstream decision making. Our approach recasts cross‑population heterogeneity into a multivariate sampling model. We propose two forecasting models for partially observed curves: a full factor analysis model that extends a general factor representation to incorporate the partially observed survival curve, and a double PCA model. The methodology is motivated by demand forecasting in short‑term rental markets, where market‑level occupancy paths can be viewed as survival curves over the booking horizon and where forecasts of future occupancy feed into dynamic pricing algorithms. We apply the models to the newly released Wheelhouse dataset, which contains time series of market occupancy curves for 500 markets from 2017 to 2022. Model performance is assessed using the integrated quadratic distance, and we compare the proposed PCA‑based methods to Holt’s linear trend model across multiple forecast horizons. The results show that the proposed models yield accurate and stable forecasts of the remaining survival trajectory and generally outperform Holt’s method, particularly at longer horizons.
\end{abstract}

\section{Introduction}
\subsection{Motivation and data examples}
A centerpiece of the e-commerce revolution has been to facilitate the functioning of markets for heterogeneous goods \citep{dedola_et_2023}.  Items in this market are unique and subject to demand by a cohort of consumers whose specific requirements are simultaneously multifaceted and vague. The same product may be offered in different contexts, which impacts the intensity of demand. Scarcity of specific inventory elicits strong substitution effects, resulting in periods of market-wide tightness. The ability to collect data, monitor demand patterns and match one context to previous experience has enabled the development of dynamic algorithms to adjust prices, thereby appropriately matching -- and even anticipating -- persistence in demand fluctuations.

The demand for short-term vacation rentals, as promoted by platforms such as Airbnb, Homeaway, vrbo, etc., offers an interesting example of this paradigm.  Clearly, the units available are highly (sometimes perfectly) heterogeneous.  Simultaneously, each unit is available for a range of dates, which have different, though related demand patterns.  Different rental markets have distinct tempos at which their inventory is diminished. 

Property management in this landscape involves balancing a diverse set of factors, especially since the inventory offered on these platforms exhibit substantially more heterogeneity than e.g. a hotel room with a king-sized bed. Underpinning almost all nuance in the decision process is the prediction of how quickly a given night (stay-date hereafter) will book under unit settings considered to be "average".  Therefore, accurately estimating the market's total demand for a given stay-date, and the trajectory of demand intensity, is the bedrock of a price adjustment protocol grounded in decision theory.

Digital platforms facilitate regular price updates throughout the period in which the item is offered. Thus, demand estimates for each stay-date ought to be continuously adjusted as observed demand intensity for the stay-date progresses and reveals potentially unanticipated excess demand.  This is especially useful when the overall amount of inventory depleted in a given context is still low, but relatively high compared to expectations. Real-time, multivariate updates of expected future demand based on the partial observation of a set of related stay-dates and markets is therefore an essential component to building a robust dynamic pricing system.

Tracking market-wide occupancy through time has an obvious relation to a population-level survival curve. In this context, the survival curve indicates the probability that a unit for a particular stay-date has not yet been booked a given number of days before the stay-date.  If the market-wide booking intensity is faster than normal in the early stages of the booking process, it seems reasonable to assume this will persist. This analysis therefore requires a statistical model for populations of survival curves and, in particular, a model for which forecasts of the curve's remainder are straightforward to elicit based on the curve's initial section.

\subsubsection{Example: Lake Tahoe during COVID-$19$}
\begin{figure}[htbp] 
    \centering
    \includegraphics[page = 1, height = 0.3\textheight, width=0.7\textwidth]{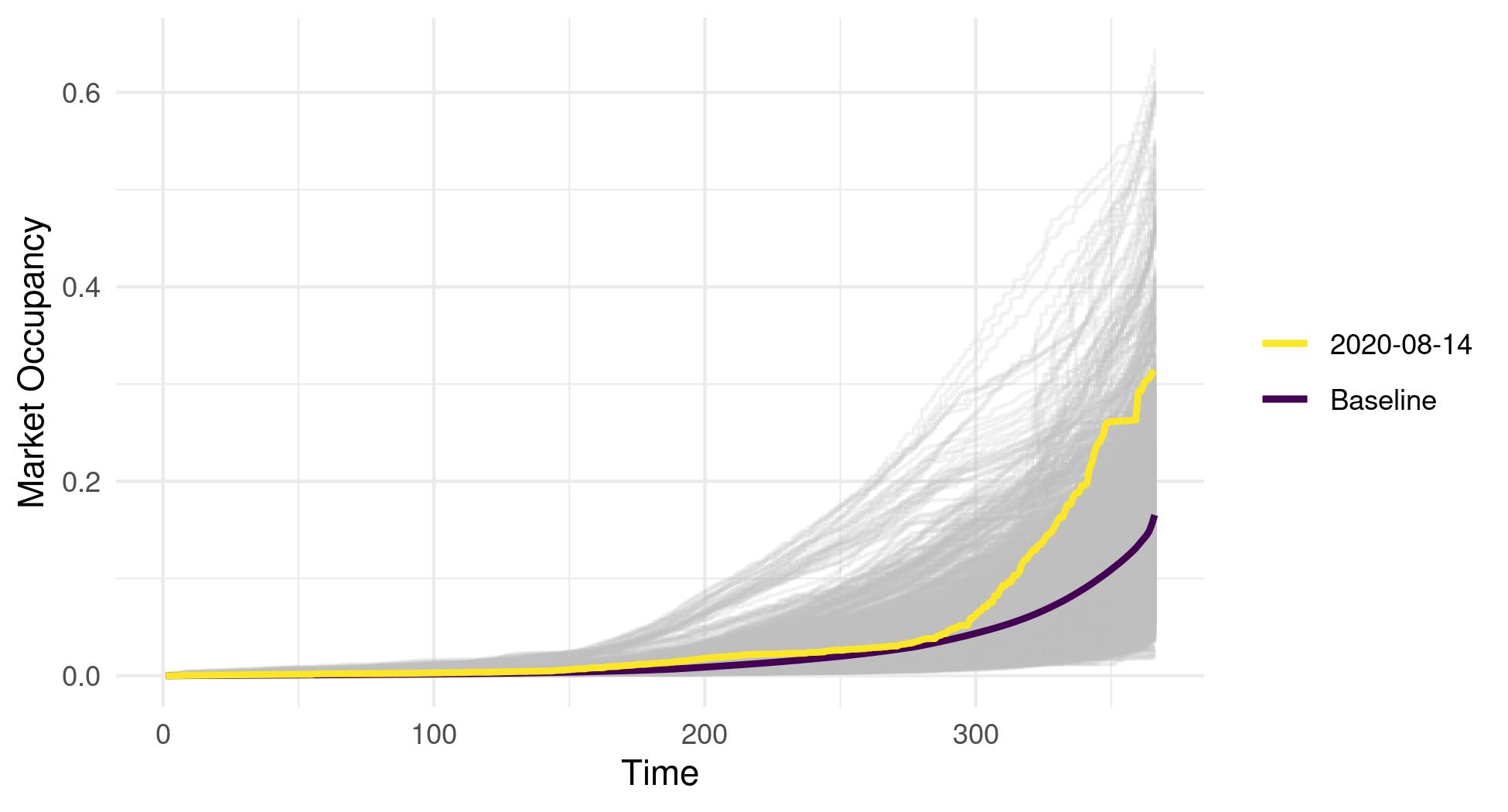} 
    \caption{Market occupancy ($1-\mathbb{S}(t)$) in Lake Tahoe. The gray lines shows all the market occupancies for all the stay dates from January $1$st, $2018$ to December $31$st, $2022$. The purple line is the baseline over the time period and the yellow line is for August $14$th $2020$.} 
    \label{fig:market_occ_LT} 
\end{figure}

Figure \ref{fig:market_occ_LT} shows examples of the market occupancy path (inverse of the survival function) for a specific market. In this figure, the gray lines are the estimated market occupancy for the Lake Tahoe region in the United States between January $1$st $2018$ and December $31$st $2022$. In addition, there are two highlighted lines. The purple line is the baseline occupancy rate over the time period, and the yellow line is a date during the COVID-$19$ pandemic. The baseline occupancy can loosely be defined as the market occupancy curve for a common stay date (the formal definition is found in Section \ref{sec:models_prediction} with a specification in Appendix \ref{sec:S_0}). Lake Tahoe is a popular vacation area both in the summer (hiking, biking, swimming) and winter (skiing). The baseline (purple line) and the yellow line track one another until $70$ days before the stay date. Then suddenly, with less virus circulating in the summer, and less restrictions, people dared to travel for summer vacation. This results in the final market occupancy ending at a much higher occupancy, even though the lines track one another up until $70$ days before the stay date.  

\subsubsection{Example: Outside Lands in San Francisco}
\begin{figure}[H] 
    \centering
    \includegraphics[page = 1, height = 0.3\textheight, width=0.7\textwidth]{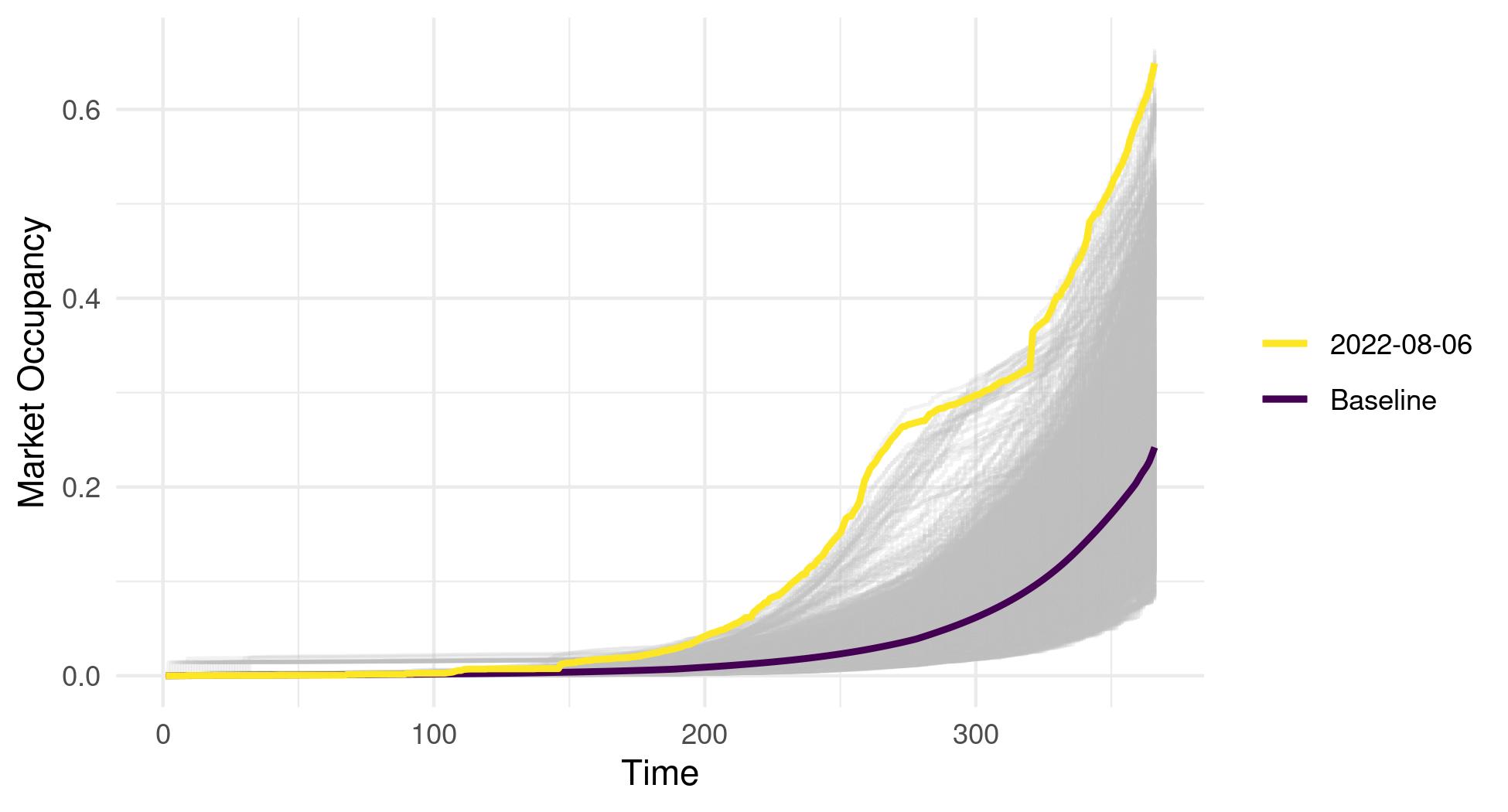} 
    \caption{Market occupancy ($1-\mathbb{S}(t)$) in San Francisco. The gray lines shows all the market occupancies for all the stay dates from January $1$st, $2018$ to December $31$st, $2022$. The purple line is the baseline over the time period and the yellow line is for $6$th of August $2022$} 
    \label{fig:market_occ_SF} 
\end{figure}
Clear deviations from the baseline are especially prevalent during market-wide events. An example of an event is Outside Lands in San Francisco, a large festival occurring every August (except during COVID-$19$). In Figure \ref{fig:market_occ_SF}, the market occupancy for $6$th of August $2022$, the second date of Outside Lands in $2022$, is highlighted in yellow. The figure shows that already $200$ days before the stay date the yellow line starts to deviate from the baseline and ends up with a much higher market occupancy in the end, indicating early bookings by individuals. 

\subsection{Literature review} \label{subsec:lit}
The short-term rental market is an important domain for forecasting models with business applications. These forecasting models often aim to maximize revenue, see for example work on call centers (see \citet{shen2008interday}), the airline industry (see \citet{lee1990airline}) and hotel demand (see \citet{haensel2011booking}, \citet{contessi2024decoding} and overview articles in \citet{weatherford2003comparison}, \citet{henriques2024hotel} and \citet{wu2025tourism}). 

Statistical models for maximizing revenue across different applications share many properties. For example, the model developed for call centers in \citet{shen2008interday} was adapted to hotel demand in \citet{haensel2011booking}. \citet{haensel2011booking} forecast the accumulated booking curve and the expected number of reservations. They combine singular value decomposition with a multivariate vector autoregressive model and additive univariate Holt-Winters model. More recently, \citet{contessi2024decoding} propose a two-step approach with PCA and an additive pickup model for forecasting hotel demand. The additive pickup model is discussed for hotels in \citet{weatherford2003comparison} and \citet{fiori2019reservation}. 

The exponential smoothing models \citep{holt1957forecasting, BrownRobertGoodell1959Sffi, winters1960forecasting}) are popular models for forecasting tourism demand. Following \citet{hyndman2018forecasting}, Holt's linear trend model \citep{holt1957forecasting} consists of a forecast equation, a level equation and a trend equation. The equations are
\begin{align}
\label{eq:Holt}
    \widehat{y}_{t+h \mid t} & = \ell_t + hb_t && \text{Forecast equation}\\
    \ell_t & = \alpha y_t + (1-\alpha)(\ell_{t-1} + b_{t-1}) && \text{Level equation}\\
    b_t & = \beta^* (\ell_t - \ell_{t-1}) + (1-\beta^*)b_{t-1} && \text{Trend equation},\label{eq:Holt_2}
\end{align}
where $\ell_t$ is the level of the series at time $t$, $b_t$ is an estimate of the slope at time $t$, $\alpha$ is the smoothing parameter for the level and $\beta^*$ is the smoothing parameter for the trend. The Holt-Winters seasonal model is an additional extension of Holt's linear trend method that also captures seasonality. Some examples of forecasting tourism demand with exponential smoothing methods are \citet{ATHANASOPOULOS200819}, \citet{ATHANASOPOULOS2009146} and \citet{ATHANASOPOULOS2011822}.  

In this paper, the methodological work focuses on forecasting the trajectory of partially observed survival curves, an application closely tied to functional data analysis, an important topic in sociological, environmental, transportation, biological and clinical research \citep{Jiao02012023}. The main focus of the research has been on completely observed curves, and especially on the functional auto-regressive model (FAR), see for example \citet{bosq2000linear} and \citet{antoniadis2003wavelet}. Further development of the methods was for example done in \citet{KARGIN20082508}, which predicts curve-valued autoregression processes, via a technique called predictive factor decomposition. Considerable work on functional data uses functional principal components analysis (FPCA). For example \citet{Aue02012015} use FPCA to transform a functional time series into a vector time series of FPCA with a lower dimension. Other relevant work on FPCA is forecasting functional time series on mortality and fertility rates (see, e.g., \citet{HYNDMAN20074942}). This model was further developed in \citet{HYNDMAN2009199}. Finally, \citet{Jiao02012023} develop a new method for predicting an unobserved part of future trajectories for functional time series data by combining information from the past daily trajectories with the partially observed trajectory of the curve. 

\subsection{Our contributions}
The main contribution of this paper is the development of two prediction models for partially observed survival curves that jointly exploit information from fully observed historical curves and the currently observed partial curve. In many applications, including our motivating example of market-level booking curves, classical time series models cannot effectively use both sources of information and tend to produce poor and unstable forecasts, especially at longer forecast horizons. Our starting point is a time-varying Type-II Lehmann model, which recasts cross-population heterogeneity in survival curves into a multivariate sampling model on transformed real-valued trajectories. On this transformed scale, we construct two PCA-based forecasting models. 

The first model is a full factor analysis model for partially observed curves, where we extend a standard factor representation of the transformed survival trajectories to incorporate the partially observed curve directly. The second model is a double PCA for partially observed curves, where we perform two separate principal component analyses on the segments of the historical curves before and after the forecast horizon and link the corresponding scores via a multivariate linear regression. To the best of our knowledge, we are the first to propose these two prediction models for time series of multiple, partially observed survival curves. Both models can be augmented with calendar effects such as day of the week and month indicators to better capture systematic seasonal patterns in the data. 

Our focus is on forecasting total market occupancy for a future stay-date in the short-term rental market based on currently observed occupancy levels.  However, the methodology discussed can be useful in a host of related business applications.  Any organization attempting to sell perishable stock is faced with determining at each point in the sales cycle the expected total amount of inventory to be demanded and the tempo of this demand intensity.  When substantial historical information on previous sales cycles is available (in, e.g., concert and airline tickets, demand for rideshare vehicles, sale of non-storable agricultural items), this methodology could serve as a useful approach to building a larger decision system.

The rest of the paper is organized as follows. In Section \ref{sec:models_prediction} we first introduce the time-varying Type-II Lehmann model and describe the two prediction models; the full factor analysis model and the double PCA model for partially observed curves. Section \ref{sec:model_evaluation} presents the use of the integrated quadratic distance for model evaluation. Section \ref{sec:n_dim} describes the methodology for selecting the model dimension based on predictive performance. Empirical results from the short-term rental market application are reported in Section \ref{sec:analysis}, and Section \ref{sec:Conclusion} discusses the findings and concludes.

\section{Methodology} \label{sec:models_prediction}
\subsection{Preliminaries and problem formulation} \label{subsec:prelim}
Let $\mathbb{S}(t)$ be a survival function estimated by the Kaplan-Meier estimator introduced in \citet{kaplan1958nonparametric}. Following the set-up in \citet{datapaper}, we assume there exist a family of survival curves, $\mc{S}$, where each curve is defined on $\mc{T}\subset \mathbb{N}$. Thus $\mathbb{S} \in \mc{S}$ is a monotonically decreasing function $\mathbb{S}:\mc{T} \to [0,1]$. In addition, there is a reference member $\mathbb{S}_0 \in \mc{S}$ which we call the baseline survival function. As the problem has been defined in a discrete space, we can write any $\mathbb{S}_i \in \mc{S}$, where $\mathbb{S}_i$ is the survival function for stay date $i$, as
\begin{equation}
    \mathbb{S}_i(t) = \prod_{r\in\mc{T}: r\leq t}\xi_{ir}
    \label{eq:surv_1}
\end{equation}
for $\xi_{ir} \in (0,1]$.

If only part of the curve $\mathbb{S}_i(t)$ is observed, Equation \eqref{eq:surv_1} can be rewritten as
\begin{equation}
\label{eq:cond_surv}
\mathbb{S}_i(t \mid \ell) = [\prod_{r\in\mc{T}: r\leq \ell}\xi_{ir}] [\prod_{r\in\mc{T}: \ell \leq r\leq t}\widehat{\xi}_{ir}]
\end{equation}
where $\ell$ represents the maximum time point of the observed data. The goal is to accurately estimate each individual $\widehat{\xi}_{ir}$.

Following the representation of $\mathbb{S}_i$ above, the baseline curve $\mathbb{S}_0$ can be formulated as $\mathbb{S}_0 = \prod_{r\in\mc{T}: r\leq t} \xi_{0r}$. Assume $\xi_{ir} = \xi_{0r}^{\exp(\gamma_{ir})}$, where $\gamma_{ir} \in \mathbb{R}$. If $\gamma_{it} = \gamma_i$ for all $t\in\mc{T}$, then $\mathbb{S}_i(t) = \mathbb{S}_0(t)^{\exp \gamma_i}$, which was first introduced in \citet{lehmann_rank_test} as a Type-II Lehmann model. This model is obviously connected to the Cox-proportional hazard model \citep{cox_prop_haz}. In \citet{lehmann_rank_test} and related work (see for example \citet{hall2013efficient}), a constant parameter $\gamma_i$ is assumed. This is too restrictive for our modeling framework, and we introduce time-varying parameters in the Type-II Lehman model. 

By the telescoping nature of $\mathbb{S}_i$, we obtain
\begin{equation}
\gamma_{it} = \log\left(\frac{\log(\mathbb{S}_{i}(t) / \mathbb{S}_{i}(t - 1))}{\log(\mathbb{S}_0(t) / \mathbb{S}_0(t - 1))}\right).
\label{eq:gamma_t}
\end{equation}
All the steps in the transformation can be found in Appendix \ref{app: background}. By studying $\gamma_{it}$ instead of $\mathbb{S}_{i}(t)$, we have reduced the problem of modeling the dynamics of $\mathbb{S}_i$ across the population $\mc{S}$ to modeling $\bs{\gamma}_i \in \mathbb{R}^{T}$. This provides a considerably more useful model space as both monotonicity and restriction to the unit interval are no longer required to be respected when modeling process dynamics and heterogeneity.

\subsection{Model 1: Full factor analysis} \label{subsec:factor}
The vectors $\bs{\gamma}_i \in \mathbb{R}^T$ are likely to be subject to substantial inter-component noise, which may not have particular use either in forecasting or in downstream decision making. A principal component analysis (PCA) \citep{hotelling1933analysis, anderson1963asymptotic} framework is therefore suitable for the forecasting problem. 

In Section \ref{subsec:prelim}, we introduced modeling the dynamics of $\mathbb{S}_i(t)$ across a population by modeling $\bs{\gamma}_i \in \mathbb{R}^T$. The vector $\bs{\gamma}_i$ can be written in a lower dimensional factor form as 
\begin{align}
\bs{\gamma}_i &= \sum_{d = 1}^D \alpha_{id}\bs{U}_d + \bs{\mu}_i + \bs{\epsilon}_i  \label{eq:factor_form}
\end{align}
where $\alpha_{id} \in \mathbb{R}$ is a set of $D$ latent variables specific to $i$ and $\bs{\mu}_i$ is a D-dimensional vector. $\bs{\alpha}_i$ is called the loading coefficient. $\bs{U}_d \in \mathbb{R}^T$ and the columns of $\bs{U}_d$ are called factor loadings, which capture the correlations between the observed variables \citep[p.~584]{bishop2006pattern}. Furthermore, $\bs{\epsilon}_i\sim \mc{N}_T(0, \bs{\Xi})$, where $\bs{\Xi}$ is a diagonal matrix, and the diagonal is a source of uncorrelated idiosyncratic noise. 

We rewrite Equation \eqref{eq:factor_form} as
\begin{align}
\tilde{\bs{\gamma}}_i&= \sum_{d = 1}^D \alpha_{id}\bs{U}_d + \bs{\epsilon}_i, \label{eq:factor_form2}
\end{align}
where $\tilde{\bs{\gamma}}_i = \bs{\gamma}_i - \bs{\mu}_i$. Following \citet[p.~584]{bishop2006pattern}, this becomes
\begin{align}
	p(\tilde{\bs{\gamma}}_i \mid \bs{\alpha}_i) = \mathcal{N}\Big(\tilde{\bs{\gamma}}_i \mid \sum_{d = 1}^D \alpha_{id}\bs{U}_d, \bs{\Xi}\Big),
	\label{eq:gamma_a}
\end{align}
where $\bs{\Xi}$ is a $D \times D$ diagonal matrix. The observed vectors $\bs{\gamma}_1, \dots, \bs{\gamma}_D$ are independent given the latent variable $\bs{\alpha}_i$. The factor analysis model explains the observed covariance structure of the data by representing the independent variance associated with each coordinate of the matrix $\bs{\Xi}$ and by capturing the covariance between the variables in the matrix $\bs{U}_d$ \citep[p.~584]{bishop2006pattern}. Following standard results for the multivariate normal distribution (see \citet{bishop2006pattern}[p.~113, 585]), and from Equation \eqref{eq:factor_form2}, this results in $p(\tilde{\bs{\gamma}}_i) = \mathcal{N}(\bs{0}, \bs{\Omega})$, where $\bs{\Omega} = \bs{U} \bs{U}^T + \bs{\Xi}$.

\subsubsection{Predicting partially observed curves} \label{subsubsec:partial_factor}
We now consider the following problem.  Let $A,B\subset \mc{T}$ form a partition of $\mc{T}$, namely that $A\cup B = \mc{T}$ and $A\cap B = \emptyset$. If $(\bs{\gamma}_i)_A$ is observed, we would like to predict the values of $(\bs{\gamma}_i)_B$. In particular, if $A$ and $B$ are sequential ($r < s$ for all $r\in A$ and $s\in B$) then this prediction problem amounts to predicting $\mathbb{S}(s)$ for all $s\in B$ upon observing the sequence $\mathbb{S}(r)$ for all time points $r\in A$.  In our application, this translates to predicting the curve of market occupancy in the period $B$ having observed the quantity of the market that has booked in the initial period $A$ for a given stay date $i$.

Since we are modeling $\tilde{\bs{\gamma}}_i$, we want to forecast the unobserved part of $\tilde{\bs{\gamma}}_i$. In particular, we use $(\tilde{\bs{\gamma}}_i)_A$ to infer $\tilde{\bs{\alpha}}_i$ and subsequently $(\tilde{\bs{\gamma}}_i)_B$, thereby enabling an estimate of $\mathbb{S}_i(s)$ for all $s\in B$. From Equation ~\eqref{eq:factor_form2} we have that $(\tilde{\bs{\gamma}}_i)_A = \bs{U}_A'\bs{\alpha}_i + (\epsilon_i)_A$ where $\bs{U}_A$ is the $|A| \times D$ submatrix of $\bs{U}$ with rows according to $A$. From rules of conditional normal distribution, we get
\begin{align}
\label{eq:gamma_b_a}
    (\tilde{\bs{\gamma}}_i)_B &\mid (\tilde{\bs{\gamma}}_i)_A \sim \mathcal{N}(\widehat{\bs{\mu}}_{B \vert A}, \widehat{\bs{\Omega}}_{B \vert A}), \\
\shortintertext{where}
    \widehat{\bs{\mu}}_{B \vert A} & = \bs{\Omega}_{BA} \bs{\Omega}_{AA}^{-1} (\tilde{\bs{\gamma}}_i)_A \label{eq:mu_b_a} \\
       \widehat{\bs{\Omega}}_{B \vert A} &= \bs{\Omega}_{BB} - \bs{\Omega}_{BA}\bs{\Omega}_{AA}^{-1}\bs{\Omega}_{AB}. \label{eq:omega_b_a}
\end{align}
Based on Equations \eqref{eq:gamma_b_a}-\eqref{eq:omega_b_a}, the remainder of the curve can be estimated. One benefit of this approach is that the covariance matrix does not need to be re-estimated for every forecast horizon.

\subsection{Model 2: Double PCA}
Following the formulation in \citet[p.~563]{bishop2006pattern}, we start with an orthonormal set of $D$-dimensional basis vectors $\{\bs{u_d}\}$, where $d = 1, \dots, D$ that satisfy $\bm{u_d}^T \bs{u_k} = \delta_{dk}$, where the basis is complete. The vector of data points can therefore be written as a linear combination of the basis vectors
\begin{align}
    \bm{\gamma_i} = \sum_{d = 1}^D \vartheta_{id}\bs{u}_d,
    \label{eq:pca-all}
\end{align}
where $\vartheta_{id}$ are the scores. 

\subsubsection{Predicting partially observed curves}
In order to take the partial observed curve into account, we perform a data stratification by dividing the data in two parts. Specifically, we want to forecast $(\bm{\gamma}_{i})_B | (\bm{\gamma}_{i})_{A}$, where $B$ is the forecast horizon $r:t$, and $A$ is the observed period $1:r-1$. The second model for forecasting partially observed curves is built on the general PCA framework presented above, in addition to a stratification of the data. In order to both take historical curves and the partial curves into account, we perform two separate PCAs and connecting the scores from the two PCAs by a multivariate regression. 

We perform two separate PCAs on $(\bm{\gamma}_{i})_{A}$ and on $(\bm{\gamma}_{i})_{B}$. Writing this in a similar form as for Equation \eqref{eq:pca-all} 
\begin{align}
(\bm{\gamma}_i)_{A} & = \sum_{d = 1}^{D_1} \alpha_{id}\bs{U}_d\label{eq:gamma_A}\\
( \bm{\gamma}_i)_{B} &= \sum_{d = 1}^{D_2} \beta_{id} \bs{V}_d,
\label{eq:gamma_B}
\end{align} 
where $\alpha_{id}$ and $\beta_{id}$ are the scores for observation $i$, and $D_1$ and $D_2$ are fixed constants. 

In order to forecast, $\alpha_{id}$ and $\beta_{id}$ must be connected, such that is is possible to predict the scores for $\beta_{id}$ for each $d$ from $1$ to $D_2$, from $\alpha_{id}$ for $d$ between $1$ and $D_1$. We propose to perform a multivariate regression on the scores
\begin{equation} \label{eq:beta_alpha}
 \beta_{id} = \bm{C}_d \bm{\alpha}_i + \epsilon_{id},
\end{equation}
where the dimension for the vector $\bm{\alpha}_i$ is $D_1$ and $\bm{C}_d$ is the vector of regression coefficients. We estimate this model on the training data, from which $\beta_{id}$ can be forecasted for $d = 1, \dots, D_2$ on the test data.

\section{Model evaluation}
\label{sec:model_evaluation}
Scoring rules are commonly used to mathematically evaluate the accuracy of probabilistic forecasts. Specifically, a scoring rule $s$ is a function $s: \mathcal{F} \times \Omega \rightarrow \mathbb{R}$ that compares a predictive distribution $F \in \mathcal{F}$ and an observed event $Y \in \Omega$ and returns a numerical score, where a lower score indicates a better forecast. Here, $\mathcal{F}$ is an appropriate class of probability distributions on the outcome space $\Omega$. The scoring rule is considered proper if the expected score is minimized when the predictive distribution $F$ agrees with the true distribution $G$ of the event $Y$. Formally, this can be written as $s(G, G) = \mathbb{E}_G s(G, Y) \leq \mathbb{E}_G s(F, Y) = s(F, G)$ for all $F, G \in \mathcal{F}$, see, e.g., \citet{Gneiting01032007} for a review. 

Each proper scoring rule $s$ is associated with a proper score divergence $d: \mathcal{F} \times \mathcal{F} \rightarrow \mathbb{R}$ given by $d(F,G) = s(F,G) - s(G,G)$, which inherits the decision-theoretic properties of $s$ since $d(F,G) = 0$ if and only if $s(F,G) = s(G,G)$. If an empirical distribution function based on $k$ observations, $\hat{G}_k$, is observed, \citet{thorarinsdottir2013using} show that every proper score divergence $d$ is $k$-proper in the sense that  
\begin{equation*}
    \mathbb E_G d(G, \hat{G}_k) \leq \mathbb E_G d(F, \hat{G}_k)
\end{equation*}
for all $F, G \in \mathcal{F}$ and all integers $k \geq 1$. 

In our setting, the observations are given as empirical survival curves $\mathbb{S} = 1 - \hat{G}$, making it natural to evaluate competing forecasts under a proper score divergence.  Specifically, we employ the integrated quadratic divergence (IQD) given by 
\begin{equation}
\label{eq:IQD}
    d_{IQ}(F, G) = \int_{-\infty}^{\infty} (F(t) - G(t))^2 dt,
\end{equation}
where we assume that $F$ and $G$ have finite first moments, see also \citet{thorarinsdottir2013using} and \citet{thorarinsdottir2020evaluation}. For predicted and observed survival curves, $\widehat{\mathbb S}$ and $\mathbb S$, respectively, the IQD becomes  
\begin{equation}
    IQD(\widehat{\mathbb{S}}, \mathbb{S}) = \int_{1}^{T} (\mathbb{S}(t) - \widehat{\mathbb{S}}(t))^2 dt  \approx \sum_{t = 1 }^{T} [\mathbb{S}(t) - \widehat{\mathbb{S}}(t)]^2 \label{eq:iqd_sum}.
\end{equation}
Furthermore, since the aim is to approximate the expected IQD over a test set $\mathbb{S}_1, \ldots, \mathbb{S}_K$, we follow common convention and report aggregated IQD values given by $\frac{1}{K} \sum_{k=1}^K IQD(\widehat{\mathbb{S}}_k, \mathbb{S}_k)$. 

The IQD is a special case of the Cram\`{e}r distance, which is a member of the $\ell_p$ family of metrics \citep{bellemare2017cramer}. The interpretation naturally follows from Equation \eqref{eq:iqd_sum}: a lower value indicates better correspondence between $\widehat{\mathbb S}$ and $\mathbb S$, and differences may be attributed to either a shift or a dispersion change \citep{Resin&2024}. Moreover, the IQD is the score divergence of the continuous ranked probability score (CRPS), see \citet{Gneiting01032007} and \citet{thorarinsdottir2013using}. The CRPS is one of the most commonly applied proper scoring rules for probabilistic forecasts of univariate real valued events due to its ability to simultaneously assess the accuracy and the calibration, or reliability of the forecast \citep{Arnold&2024}.

\section{Dimension selection} \label{sec:n_dim}
Selecting a reasonable number of principal components can be an important parameter for both forecasting models introduced in Section \ref{sec:models_prediction}. Various rules and methods have been discussed, but most of them are viewed as ad-hoc \citep[p.~111]{JolliffeI.T.2002PCA}. For example, selecting the number of principal components that contribute $80 \%$ or $90\%$ of the total variation \citep[p.~112]{JolliffeI.T.2002PCA}. Another example is to use Kaiser's rule, which says that any principal components with a variance less than $1$ should be discarded \citep[p.~114]{JolliffeI.T.2002PCA}. The third ad-hoc method is to use a scree graph of the correlation matrix \citep[p.~116]{JolliffeI.T.2002PCA}.

\citet{Jiao02012023} argue that selecting the number of principal components that contribute, for example, $80\%$ of the total variation is often not appropriate for prediction purposes. They propose minimization of the mean squared error (MSE) of a prediction when selecting the dimension. In this paper, we propose a similar approach based on minimizing the IQD.  

We start by performing an $l$-fold cross-validation on parts of the training data (that will not be used for further training). For each fold, we calculate IQD on the test-set resulting in one IQD-value per observation. After the cross-validation, the mean using $n$ principal components is calculated for each of the test sets. Specifically
\begin{equation*}
    IQD_n = \frac{1}{K} \sum_{k = 1}^K IQD(\widehat{\mathbb{S}}_k^n,  \mathbb{S}_k^n)
\end{equation*}
for $K$ number of observations and for $n$ principal components. 

The gain from adding an additional principal component is calculated via the IQD-difference $IQD_{diff}^{n-1} = IQD_{n-1} - IQD_n$. The standard deviation is estimated via a nonparametric bootstrap \citep{efron_bootstrap} on the IQD-differences. Specifically, we sample the IQD-differences with replacement and calculate the mean. After this is performed $5000$ times, we calculate the standard deviation. To find whether there is a gain in adding an additional principal component, we perform a nonparametric Monte Carlo test. The null hypothesis, $H_0$, is that the distribution of $IQD_{diff}^{n-1}$ is centered at $0$, i.e. adding an additional principal component does not systematically improve or decrease the performance according to the IQD-value. The alternative hypothesis, $H_1$, is that including an additional principal component systematically changes the IQD-value in any direction.

\section{Results for the short-term rental market}\label{sec:analysis}
Our application is to predict market occupancy in the short-term rental market using the newly released dataset in \citet{datapaper}. The data consists of time series of market occupancy paths, or survival curves, for $500$ markets in $41$ countries worldwide from $2017$ throughout $2022$. Since the data from $2017$ appear to have a higher level of noise than the remaining years, our analysis focuses on the data from $2018-2022$. For this time period, the dataset consists of $911\,820$ individual survival curves, and since each survival curve consists of $366$ points, the dataset thus comprises $333\,726\,120$ data points. The models are re-estimated for each forecast horizon, market and forecast day results in higher computational cost that scales with these elements.

For the analysis, the data are split three ways along the temporal dimension, with the data from all 500 markets included in each subset. We use the data from $2018$ to perform dimension selection for double PCA and the full factor analysis models. Then, we train the double PCA and full factor analysis models separately for each market on data from January $1$st $2019$ to $365$ days before a given stay date, unless otherwise specified. Finally, the performance of the models is assessed over the test period from January $1$st $2022$ to December $31$st $2022$. Consequently, approximately $20\%$ of the data is used for dimension selection, up to $60\%$ of the data is used for training and testing is performed on the remaining $20\%$ of the data. 

We perform a comparison of the full factor analysis model defined by Equations \eqref{eq:gamma_b_a} - \eqref{eq:omega_b_a} and the double PCA model defined by Equations \eqref{eq:gamma_A} - \eqref{eq:beta_alpha}. The market occupancy paths are expected to exhibit seasonal variations as well as day-of-the-week patterns. We therefore investigate stratifying the training data along these dimensions, to issue forecasts that take such variations into account. This yields a total of four PCA-based models. When we include day of the week and month for double PCA and full factor analysis, the training period is from January $1$st $2019$ to the number of forecasting days before the specific stay date, a consequence of a reduced quantity of data.

We further compare the PCA-based models to Holt's linear trend model, the classical time series model presented in Section \ref{subsec:lit}. For Holt's linear trend model, the parameters $\alpha$ (smoothing for level) and $\beta^*$ (smoothing for trend) from Equations \eqref{eq:Holt} - \eqref{eq:Holt_2} are estimated automatically in the $\textit{holt}$-method from the $\textit{forecast}$-package created by \citet{forecast_package}. The model is, for each forecast instance, trained on the available partially observed curve which evolution is to be forecasted. 

The remainder of the section is organized as follows. We first describe the data stratification to include day-of-the-week and seasonal effects in the models, as well as results for selecting the dimension of the PCA-based models. We then present the overall results for all $500$ markets, followed by specific examples that demonstrate the differences in market behavior. Note that individual results are here presented as market occupancy curves (inverse of the survival curve) for ease of interpretability.    

\subsection{Data stratification} \label{sec: data_strat}
\begin{figure}[H]
    \begin{subfigure}{0.49\linewidth}
        \includegraphics[page = 1, height=0.20\textheight, width=\linewidth]{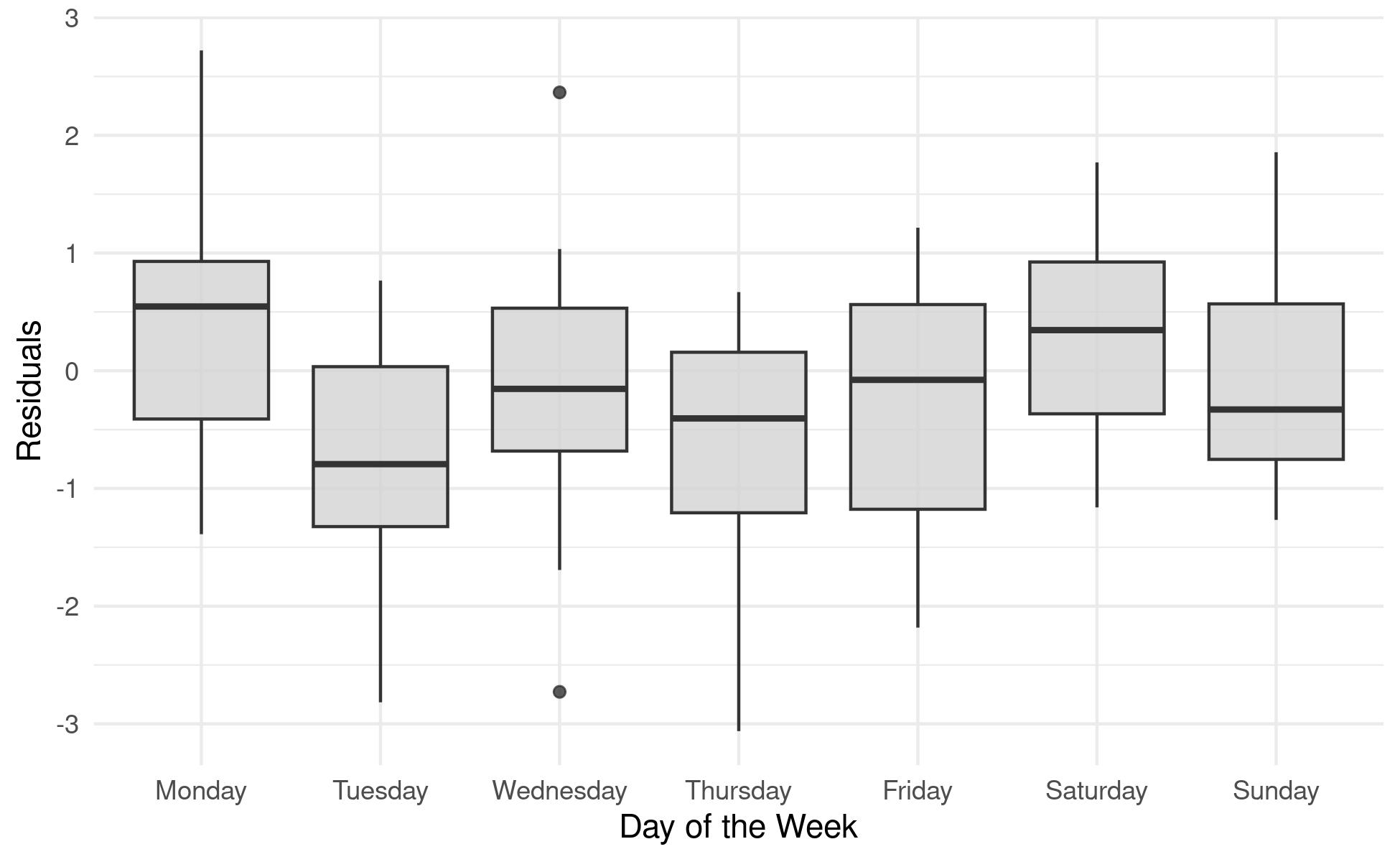}
        \caption{August residuals grouped by day of the week.}
        \label{fig:dotw_1}
    \end{subfigure}
    \hfill
    \begin{subfigure}{0.49\linewidth}
        \includegraphics[page = 1, height=0.20\textheight, width=\linewidth]{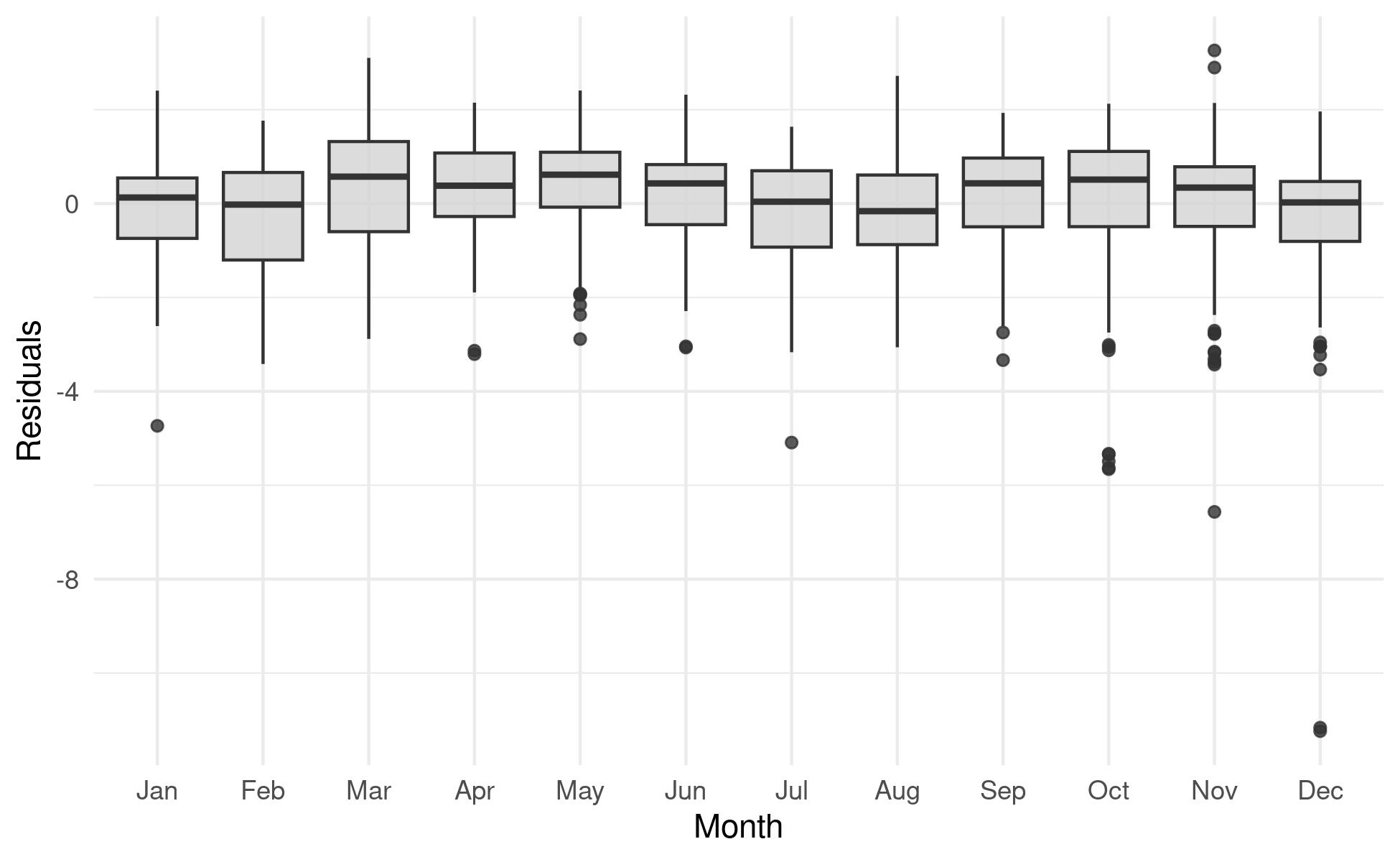}
        \caption{Residuals grouped by month.}
        \label{fig:months_1}
    \end{subfigure}
    \caption{In-sample residuals over the training period $2019-2022$ for double PCA in San Francisco, CA. The figure shows the residuals for the market occupancy $60$ days before the stay date for a forecast horizon of $90$ days,}
    \label{fig:residual_plots_market1}
\end{figure}

\begin{figure}[H]
    \begin{subfigure}{0.49\linewidth}
        \includegraphics[page = 1, height=0.20\textheight, width=\linewidth]{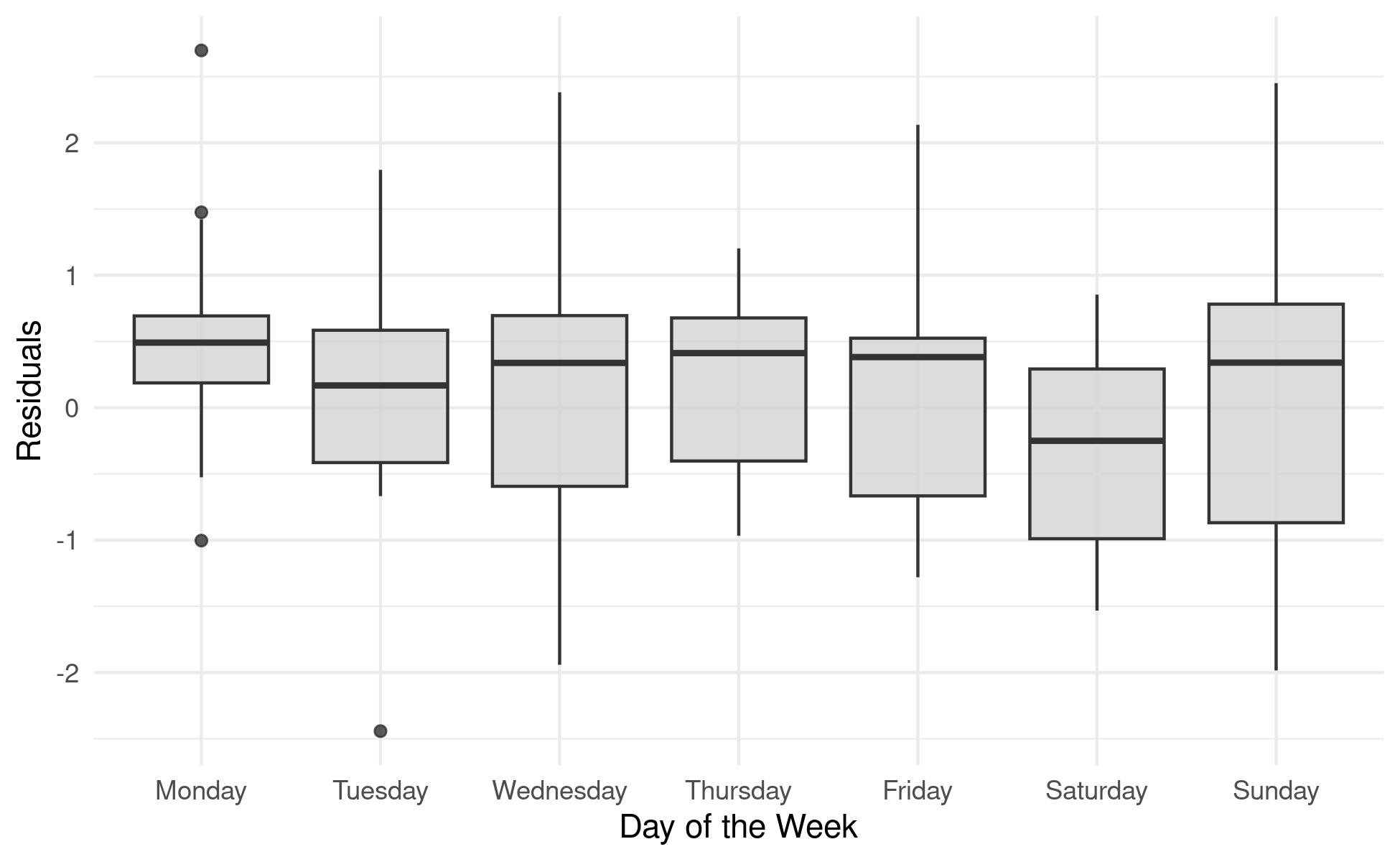}
        \caption{August residuals grouped by day of the week.}
        \label{fig:dotw_12}
    \end{subfigure}
    \hfill
    \begin{subfigure}{0.49\linewidth}
        \includegraphics[page = 1, height=0.20\textheight, width=\linewidth]{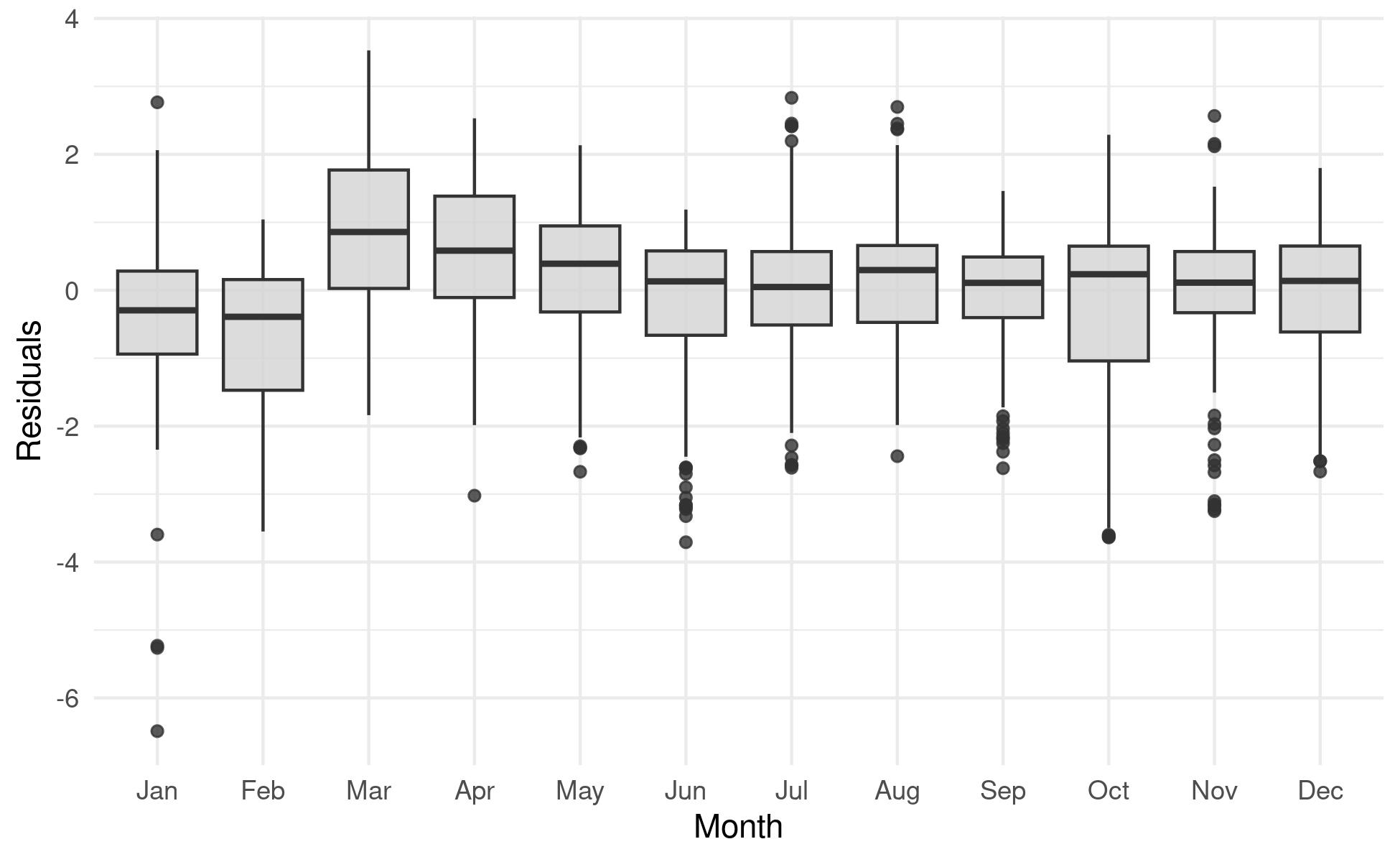}
        \caption{Residuals grouped by month.}
        \label{fig:months_12}
    \end{subfigure}
    \caption{In-sample residuals over the training period $2019-2022$ for double PCA in Destin, FL. The figures show the residuals for the market occupancy $60$ days before the stay date for a forecast horizon of $90$ days.}
    \label{fig:residual_plots_market12}
\end{figure}

To assess seasonality and day-of-the-week effects in the forecasts, we investigate the in-sample residual patterns for the training set when the residuals are stratified along these dimensions. Examples of the results of such investigations for the double PCA model are shown in Figures \ref{fig:residual_plots_market1} and \ref{fig:residual_plots_market12} for two markets. Figure \ref{fig:residual_plots_market1} focuses on San Francisco, CA which is a typical city market, while Figure \ref{fig:residual_plots_market12} shows the corresponding results for Destin, FL, a popular sunny vacation destination. In both figures, we observe strong patterns in the residuals. For example, Figure \ref{fig:dotw_1} shows a clear difference in the residuals between a Tuesday and a Saturday in August in San Francisco. Similarly, we observe stark differences between February and March residuals in Destin as displayed in Figure \ref{fig:months_12}. Further, the residual patterns differ substantially between the two locations. 

Based on the analysis exemplified in Figures \ref{fig:residual_plots_market1} and \ref{fig:residual_plots_market12}, we therefore consider two versions of each of the double PCA and the full factor analysis models where the two versions differ in terms of the training data underlying a particular forecast. Specifically, we consider a model version where all data from January 1, 2019 until 365 days prior to the stay date in question in used in training the model. Secondly, we consider a training dataset stratified by both day of the week and month in the same time period. For example, if the forecast date is a Monday in September, the training data only consist of previous Mondays in September. In both cases, only data from the specific market is used to train the models, and all parameters, including the factor loadings, are re-estimated for each forecast instance. 

\subsection{Dimension selection} \label{subsec:dim_emp}
To select the number of principal components used in the four PCA-models, we perform a $10$-fold cross-validation on the data from January $1$st $2018$ to December $31$st $2018$ for all $500$ markets and six different forecast horizons. In these investigations, we restrict the number of principal components between $1$ and $20$. For the double PCA models, the dimensions in the training and test period of each curve are assumed equal, that is, $D_1=D_2$ in Equations \eqref{eq:gamma_A} and \eqref{eq:gamma_B}. The overall results are given in Table \ref{tab:n_pc}. The results show that the full factor model requires fewer principal components than the double PCA model, with the optimal number of principal components for the full factor model ranging from $2$ to $5$. For the double PCA model, however, the optimal dimensionality varies between $3$ and $11$, with data stratification affecting the results for forecast horizons of $120$ days or longer. See Appendix \ref{app:dimension selection} for further details. 

\begin{center}
\captionof{table}{Number of principal components used for the four PCA-based models. If \ding{51}, day of the week and month effects are included in the model. For double PCA, same dimensionality is used to represent the training and the test period of each curve.}\label{tab:n_pc}
\begin{tabular}{l c c c c c c c} 
 \toprule
  & & \multicolumn{6}{c}{Forecast horizon} \\
\cline{3-8} 
 Model & D/M & $30$ & $60$ & $90$ & $120$ & $150$ & $200$\\ 
 \toprule
 %\addlinespace[0.5em]
 Full factor & \ding{55} & $4$ & $4$ & $4$ & $3$ & $2$ & $5$\\
 %\addlinespace[0.5em]
  Full factor & \ding{51} & $4$ & $4$ & $4$ & $3$ & $2$ & $5$\\
 %\addlinespace[0.5em]
  Double PCA & \ding{55} & 5 & 7 & 7 & 11 & 5 & 7\\
 %\addlinespace[0.5em]
 Double PCA & \ding{51} & 5 & 7 & 7 & 6 & 5\footnotemark & 3\footnotemark \\

 \bottomrule
\end{tabular}
\end{center}
\footnotetext[1]{For market $422$ and $440$ this number is 3.}
\footnotetext[2]{For market $440$ this number is 1.}

\subsection{Overall predictive performance} \label{subsec:overall}
To assess the overall predictive performance, we use the dimensions listed in Table \ref{tab:n_pc} for the four PCA-based models to forecast the partially observed survival curves in the test period from January $1$st $2022$ to December $31$st $2022$ and compare against Holt's linear trend model. The methods are assessed over all $500$ markets for six different forecast horizons: $30$, $60$, $90$, $120$, $150$ and $200$ days. 

\begin{center}
\captionof{table}{IQD-values, aggregated over all markets and stay dates from January $1$st $2022$ to December $31$st $2022$, for $6$ different forecast horizons. The best score for each forecast horizon is indicated in bold. If \ding{51}, day of the week and month effects are included in the model.}\label{tab:overall_perf}
\begin{tabular}{l c P{1cm} P{1cm} P{1cm} P{1cm} P{1cm} P{1cm}} 
 \toprule
 & & \multicolumn{6}{c}{Forecast horizon} \\
\cline{3-8} 
Model & D/M  & $30$ & $60$ & $90$ & $120$ & $150$ & $200$\\ 
 \midrule
 Full factor & \ding{55} &  $0.063$ & $0.186$ & $0.356$ & $0.604$ &$0.935$ & $5.518$\\
  Full factor & \ding{51} &$0.066$ & $0.184$ & $0.318$ & $0.453$ & $0.599$ & $1.582$\\
  Double PCA & \ding{55} &$0.063$ & $\bf 0.172$ & $\bf 0.291$  & $\bf 0.414$ &$\bf 0.568$ &  $ 0.781$\\
 Double PCA & \ding{51} & $0.066$ & $0.180$ & $0.309$ & $0.436$ & $0.585$ & $\bf 0.752 \footnotemark$\\
Holt & \ding{55} & $\bf 0.046$ & $0.220$ & $0.647$ & $1.362$ & $2.352$ & $5.200$\\
 \bottomrule
\end{tabular}
\end{center}
\footnotetext{Market $422$ is excluded since the model estimation failed to converge due to data deficiency.}

The aggregated IQD-values over all stay dates and markets for the five forecasting models are shown in Table \ref{tab:overall_perf}. At a forecast horizon of $30$ days, Holt's linear trend model outperforms the PCA-based models, while it has the worst performance for all other forecast horizons. For forecast horizons of 60 to 200 days, the double PCA model outperforms the other models. The parsimonious model,  where a single set of parameters is estimated for each market, is the top performing model for forecast horizons of 60 to 150 days, while the more highly parameterized model with calendar effects performs best for the longest forecast horizon of 200 days. 

Another notably difference is the stability of the predictions. While Holt's linear trend model ranges from an IQD value of approximately $0.05$ when the forecast horizon is $30$ days to $5.2$ when the forecast horizon is $200$ days, the double PCA model ranges from about $0.06$ to $0.8$ for the same horizons. The double PCA model can thus even $200$ days before the stay date relatively accurately predict the remaining trajectory of the booking curve. In practice, the host thus gets information at an early stage whether a given stay date is a high or low demand day. They can then accordingly increase or decrease prices. Since we forecast the entire booking trajectory, the forecast may also give an indication of the time points it is beneficial to increase or decrease the price.

The final $30$ days of the forecast horizon are crucial for determining the final booking probability, as well as for other downstream decisions such as pricing. Table \ref{tab:overall_perf_last_30} presents an assessment of the predictive performance over the final $30$ final only, regardless of the forecast horizon. Comparing Table \ref{tab:overall_perf} and \ref{tab:overall_perf_last_30}, we see that the last $30$ days are key determinants of the overall predictive performance. For example, for a forecast horizon of $200$ days, the score values for the double PCA models in Table \ref{tab:overall_perf_last_30} constitute roughly $60\%$ of the score values reported in Table \ref{tab:overall_perf}, while the last $30$ days comprise $15\%$ of the forecast horizon of $200$ days. Another central finding evident from Table \ref{tab:overall_perf_last_30} is that for a forecast horizon of $150$ days, the model that yields the lowest overall IQD-value is different than that for the final $30$ days only. Specifically, for the final $30$ days, it is beneficial to include day of the week and month effects.

\begin{center}
\captionof{table}{Mean IQD-values, aggregated over all markets and stay dates from January $1$st $2022$ to December $31$st $2022$, for the final $30$ days of each forecast horizon. The best score for each forecast horizon is indicated in bold. If \ding{51}, day of the week and month effects are included in the model. }\label{tab:overall_perf_last_30}
\begin{tabular}{l c P{1cm} P{1cm} P{1cm} P{1cm} P{1cm} P{1cm}} 
 \toprule
 & & \multicolumn{6}{c}{Forecast horizon} \\
\cline{3-8} 
Model & D/M  & $30$ & $60$ & $90$ & $120$ & $150$ & $200$\\ 
 \midrule
 Full factor & \ding{55} & $0.063$ &$0.173$ & $0.286$ & $0.413$ &$0.538$ & $1.574$\\
  Full factor & \ding{51} & $0.066$ & $0.171$ & $0.260$ & $0.332$ & $0.394$ & $0.627$\\
  Double PCA & \ding{55} & $0.063$ &  $\bf 0.161$ & $\bf 0.245$ & $\bf 0.317$ & $ 0.397$ & $0.491$\\
 Double PCA & \ding{51} &$0.066$ & $0.168$ & $0.254$ & $0.324$ & $\bf 0.391$ & $\bf 0.459\footnotemark$  \\
Holt & \ding{55} & $\bf 0.046$ & $0.207$ & $0.549$ &$1.050$ & $1.612$ & $2.553$\\
 \bottomrule
\end{tabular}
\end{center}

\begin{figure}[H]
    \begin{subfigure}{0.49\linewidth}
        \includegraphics[page = 1, height= 0.20\textheight, width=\linewidth]{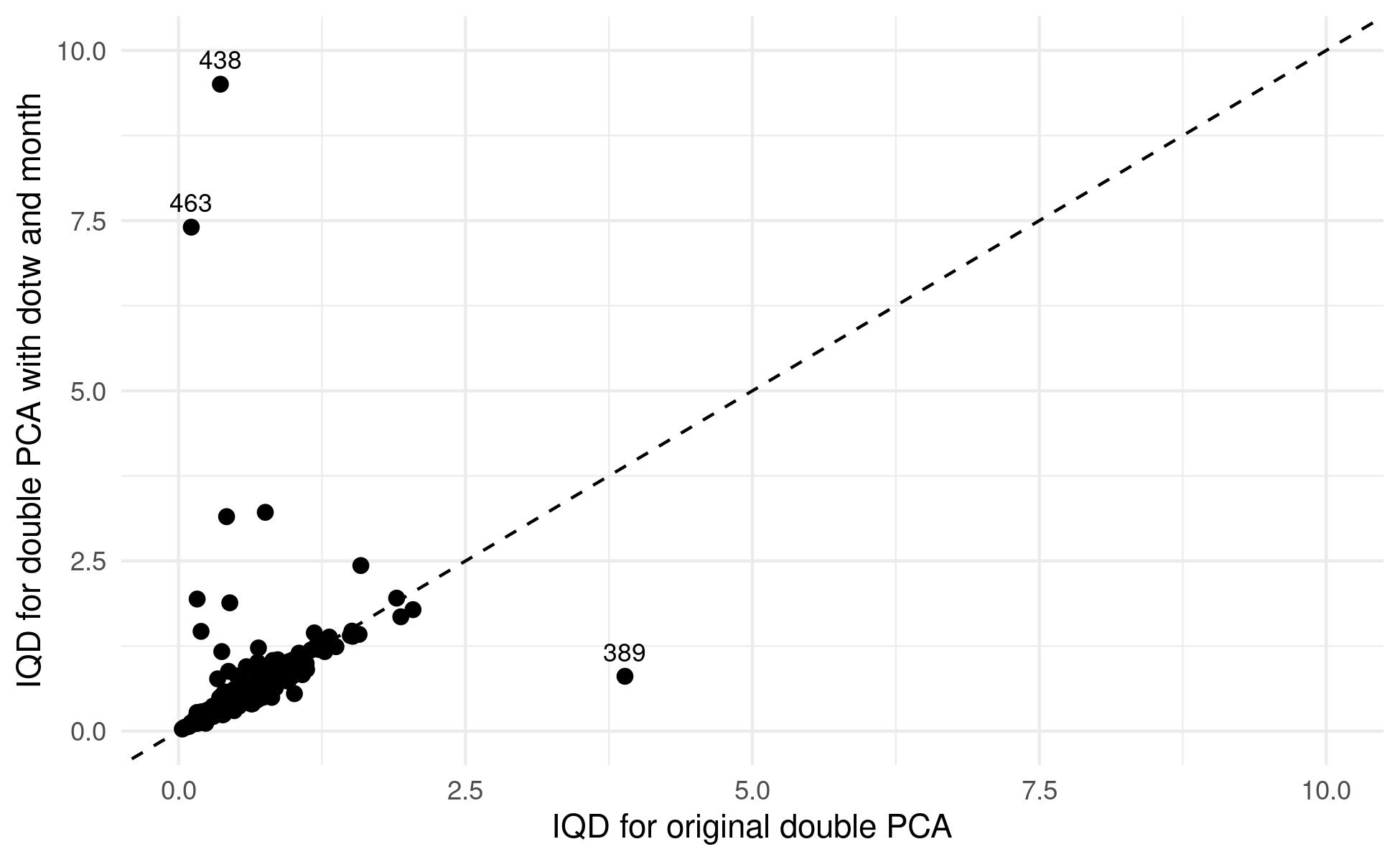}
        \caption{Forecast horizon of $150$ days.}
        \label{fig:IQD_150_a}
    \end{subfigure}
    \hfill
    \begin{subfigure}{0.49\linewidth}
        \includegraphics[page = 1, height= 0.20\textheight, width=\linewidth]{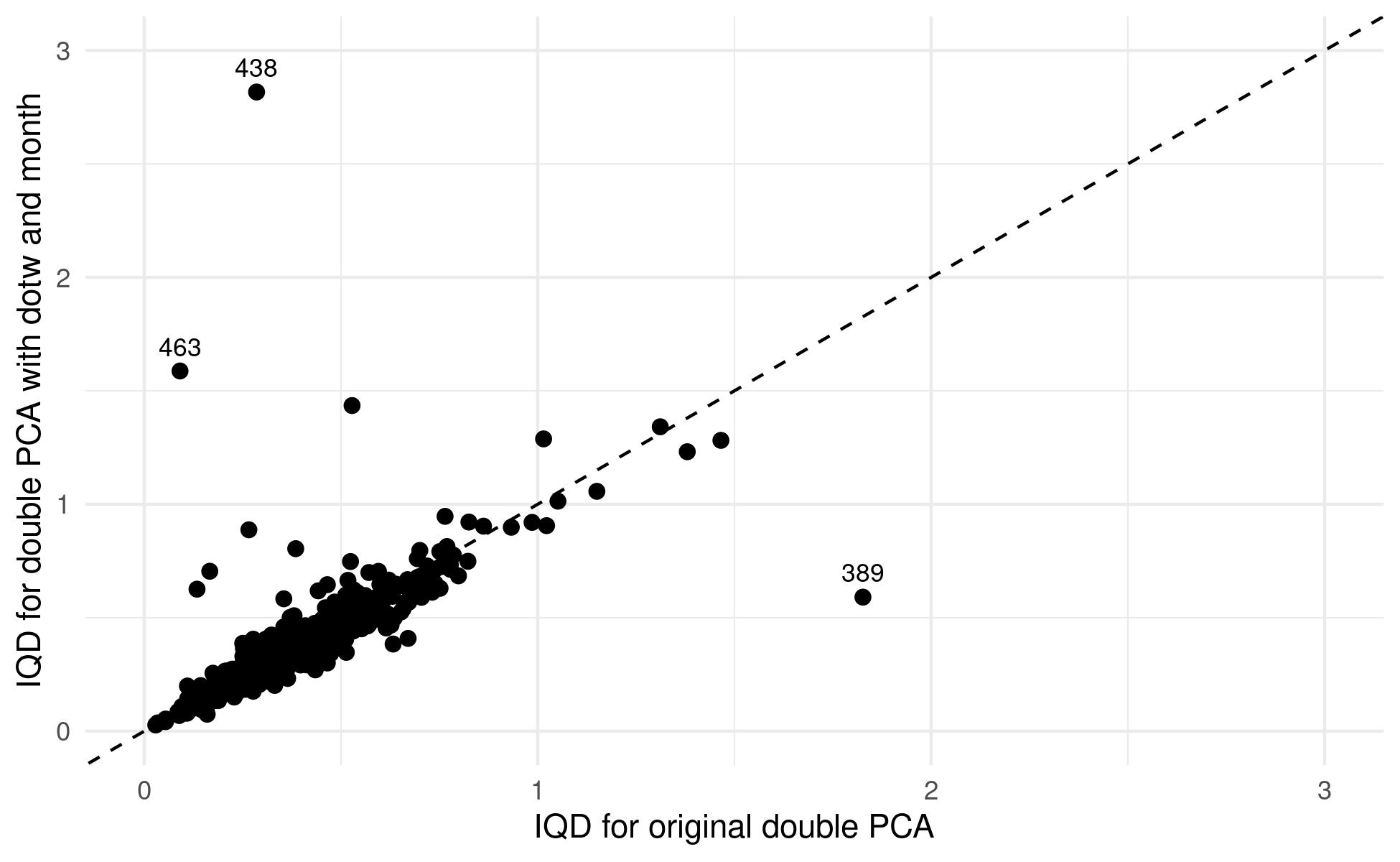}
        \caption{Final $30$ days of a forecast horizon of $150$ days.}
        \label{fig:IQD_150_b}
    \end{subfigure}
    \caption{IQD-values per market, aggregated over all stay dates in $2022$ for a forecast horizon of $150$ days, comparing the double PCA method with and without calendar effects for the full forecast horizon (a) and when focusing on the last $30$ days of the forecast horizon (b). Market identifiers are provided for three outliers.}
    \label{fig:IQD_150}
\end{figure}

\footnotetext{Market $422$ is excluded since the model estimation failed to converge due to data deficiency.}
For a further comparison of the two double PCA models for a forecast horizon of $150$ days, Figure \ref{fig:IQD_150} breaks the aggregated values reported in Tables \ref{tab:overall_perf} and \ref{tab:overall_perf_last_30} into market-specific score values. Overall, the two models yield similar results, with the model that includes calendar effects slightly outperforming the more parsimonious model at a majority of the markets. However, the score distribution for the model with calendar effects also has two substantial outliers, markets $438$ (Kalibo, Philippines) and $463$ (Eastern Antioquia, Colombia). In Kalibo, the model fails at capturing the booking patterns on roughly $50$ days out of the $365$ days in the test set, with the failure reflected across the entire forecast horizon. Similar results are observed for Eastern Antioguia, but for a specific three week period in April 2022, rather than interdispersed across the year as observed in Kalibo. Similarly, the score distribution for the parsimonious model without calendar effects has one outlier, market $389$ (Madrid, Spain). Here, the high mean score is due to the model failing to capture the market booking trends for roughly $50$ days out of the year, while the deviations are less extreme than those observed for the model with calendar effects in Kalibo and Eastern Antioguia. 

\subsubsection{Performance across markets}

\begin{figure}[H]
  \centering

  \begin{subfigure}{\linewidth}
    \centering
    \includegraphics[page = 1, height=0.35\textwidth, width=0.9\linewidth]{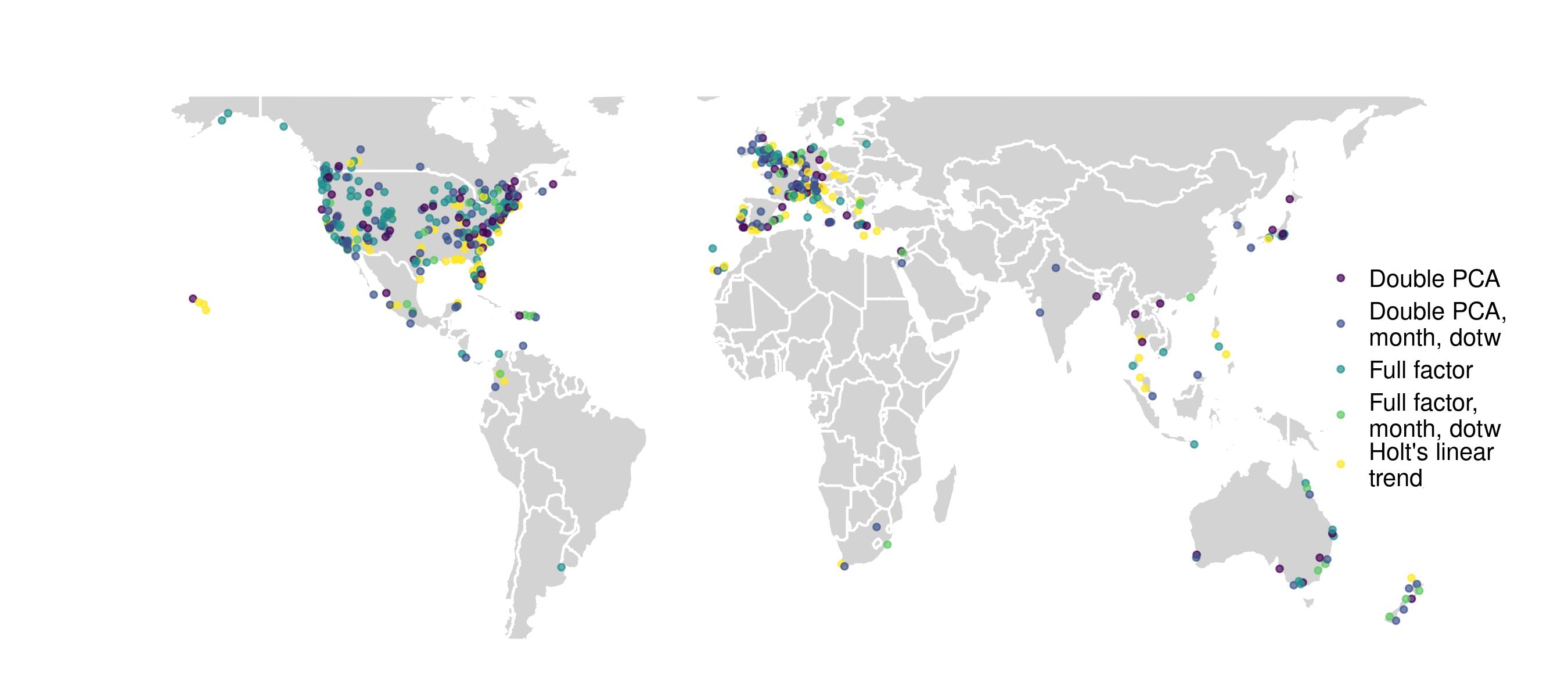}
    \caption{Forecast horizon of $60$ days.}
    \label{fig:map_60_best_iqd}
  \end{subfigure}

  \vspace{0.8em}

  \begin{subfigure}{\linewidth}
  \centering
  \includegraphics[page = 1, height=0.35\textwidth, width=0.9\linewidth]{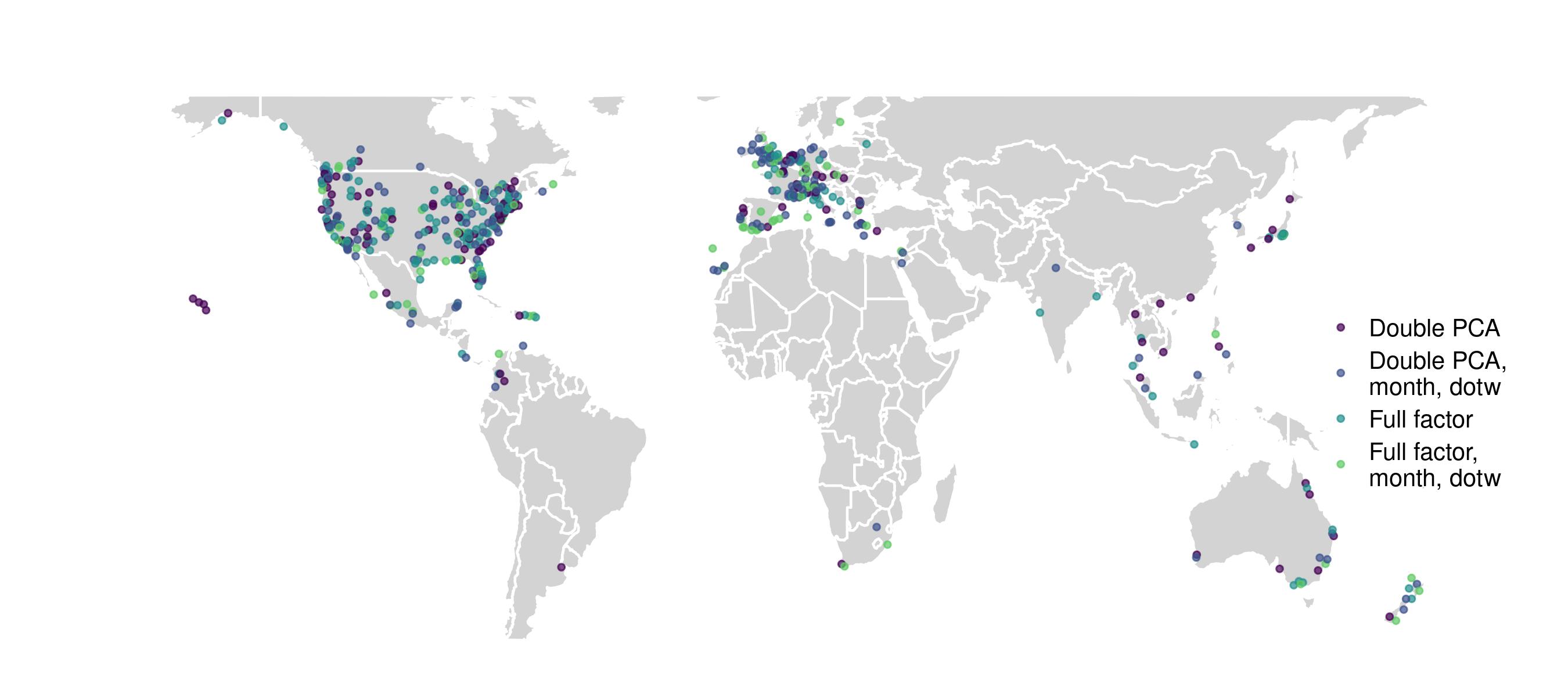}
        \caption{Forecast horizon of $120$ days.}
        \label{fig:map_120_best_iqd}
  \end{subfigure}

  \caption{Best performing method as measured by the IQD aggregated over all stay dates in $2022$ for all $500$ markets and two forecast horizons.}
  \label{fig:best_iqd}
\end{figure}

To compare the results across markets, Figure \ref{fig:best_iqd} shows the best performing method for each market for forecast horizons of $60$ and $120$ days. At a forecast horizon of $60$ days, all five models are represented at least once, cf. Figure \ref{fig:map_60_best_iqd}. Furthermore, some geographical structures may be noticed. For example, Holt's linear trend model never performs best on the West Coast of the US. 

At a forecast horizon of $120$ days, Holt's linear trend model is universally inferior, while the four PCA-based models all perform best in at least one market, cf. Figure \ref{fig:map_120_best_iqd}. While some clusters are present, such as the double PCA model without calendar effects ranking highest in all four markets in Hawaii, while the double PCA with calendar effects generally performs well in the United Kingdom, no overall structures are apparent.

\begin{figure}[H]
    \begin{subfigure}{0.49\linewidth}
        \includegraphics[page = 1, height= 0.20\textheight, width=\linewidth]{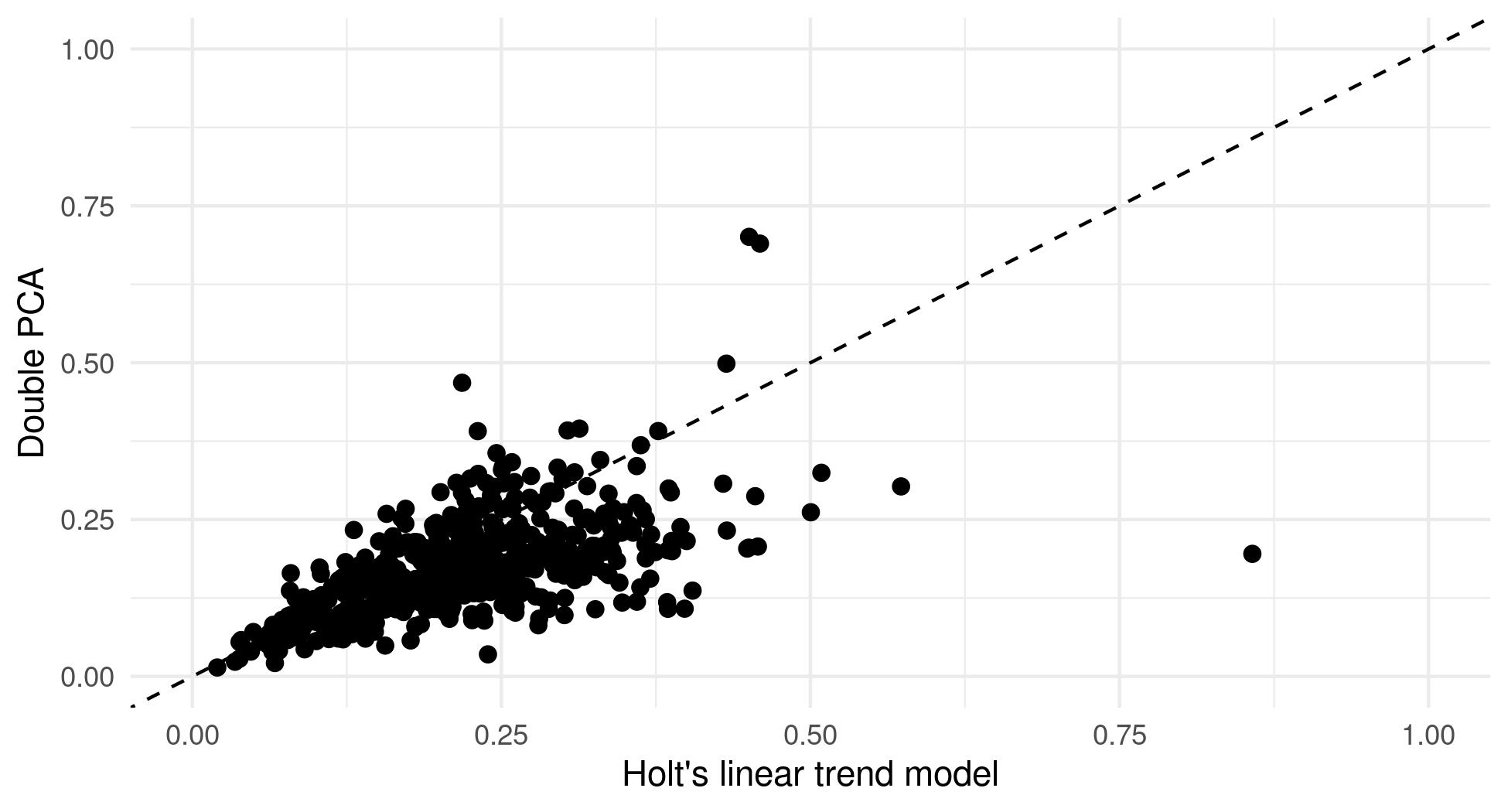}
        \caption{Forecast horizon of $60$ days.}
        \label{fig:holt_double_pca_comp_60}
    \end{subfigure}
    \hfill
    \begin{subfigure}{0.49\linewidth}
        \includegraphics[page = 1, height= 0.20\textheight, width=\linewidth]{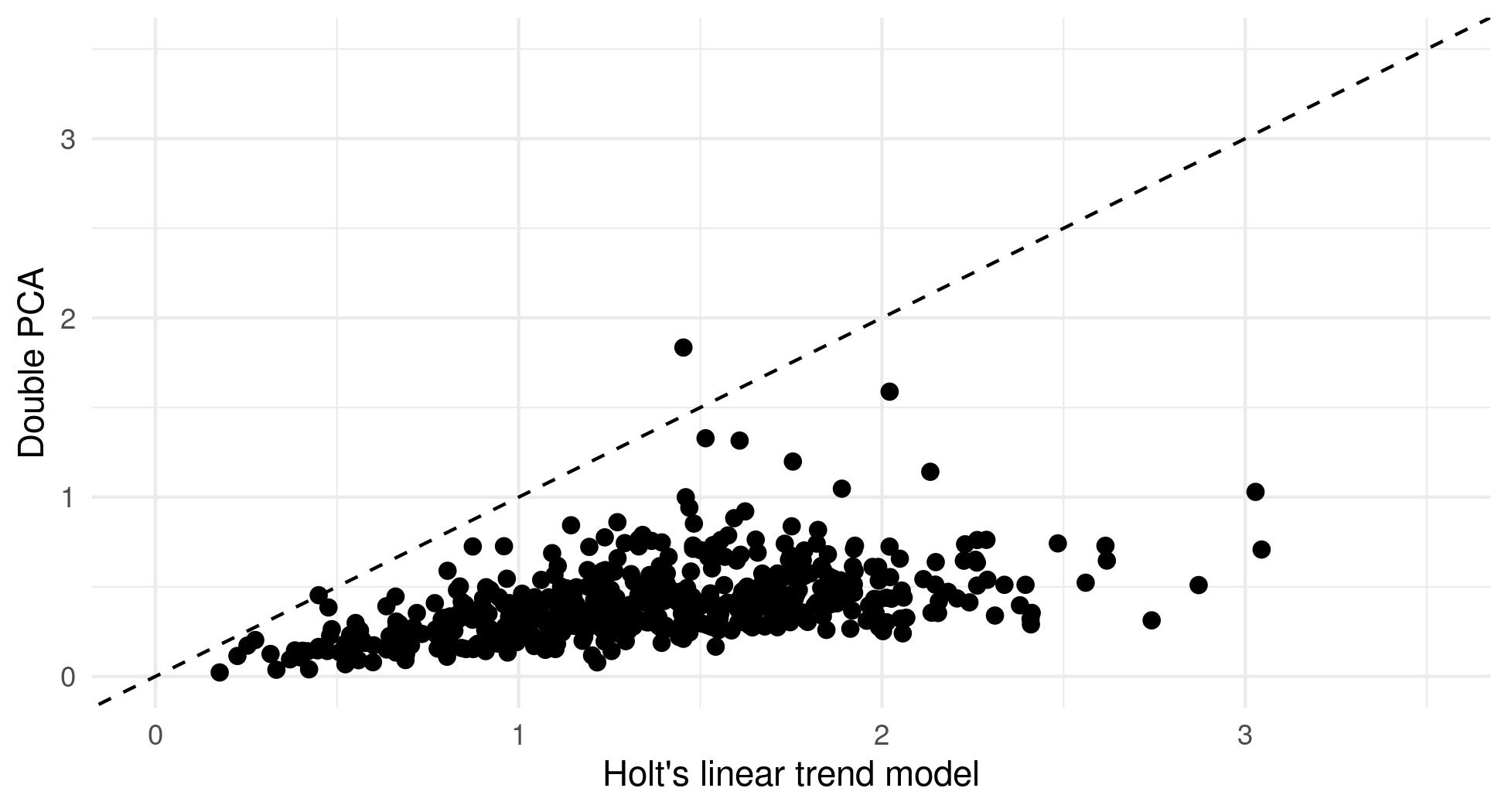}
        \caption{Forecast horizon of $120$ days.}
        \label{fig:holt_double_pca_comp_120}
    \end{subfigure}
    \caption{Market-specific IQD-values aggregated over all stay dates in $2022$ under Holt's linear trend model ($x$-axes) compared to double PCA ($y$-axes) for two forecast horizons.}
    \label{fig:holt_double_pca_comp}
\end{figure}

The differences in performance across markets between Holt's linear trend model and the parsimonious double PCA model without calendar effects are further highlighted in Figure \ref{fig:holt_double_pca_comp}. At a forecast horizon of $60$ days, cf. Figure \ref{fig:holt_double_pca_comp_60}, the points are spread around the diagonal. Majority of the points fall below the diagonal, indicating that Holt's linear trend model has higher IQD-values than double PCA for a majority of markets. For a forecast horizon of $120$ days, however, only one point is distinctly above the diagonal, cf. Figure \ref{fig:holt_double_pca_comp_120}.  

There are at least two reasons for the decreasing performance of Holt's linear trend model with increasing forecast horizon. Firstly, the model only takes the partially observed current curve into account, yielding less training data for longer forecast horizons. Secondly, the curve estimate is highly sensitive to the slope of the partially observed true curve and the market occupancy curves often start out almost flat, so that no information about the later uptick in bookings is available in the training data. 

\subsection{Case studies} \label{subsec:case_studies}
\begin{figure}[H] 
    \centering
    \includegraphics[page = 1, height = 0.3\textheight, width=0.7\textwidth]{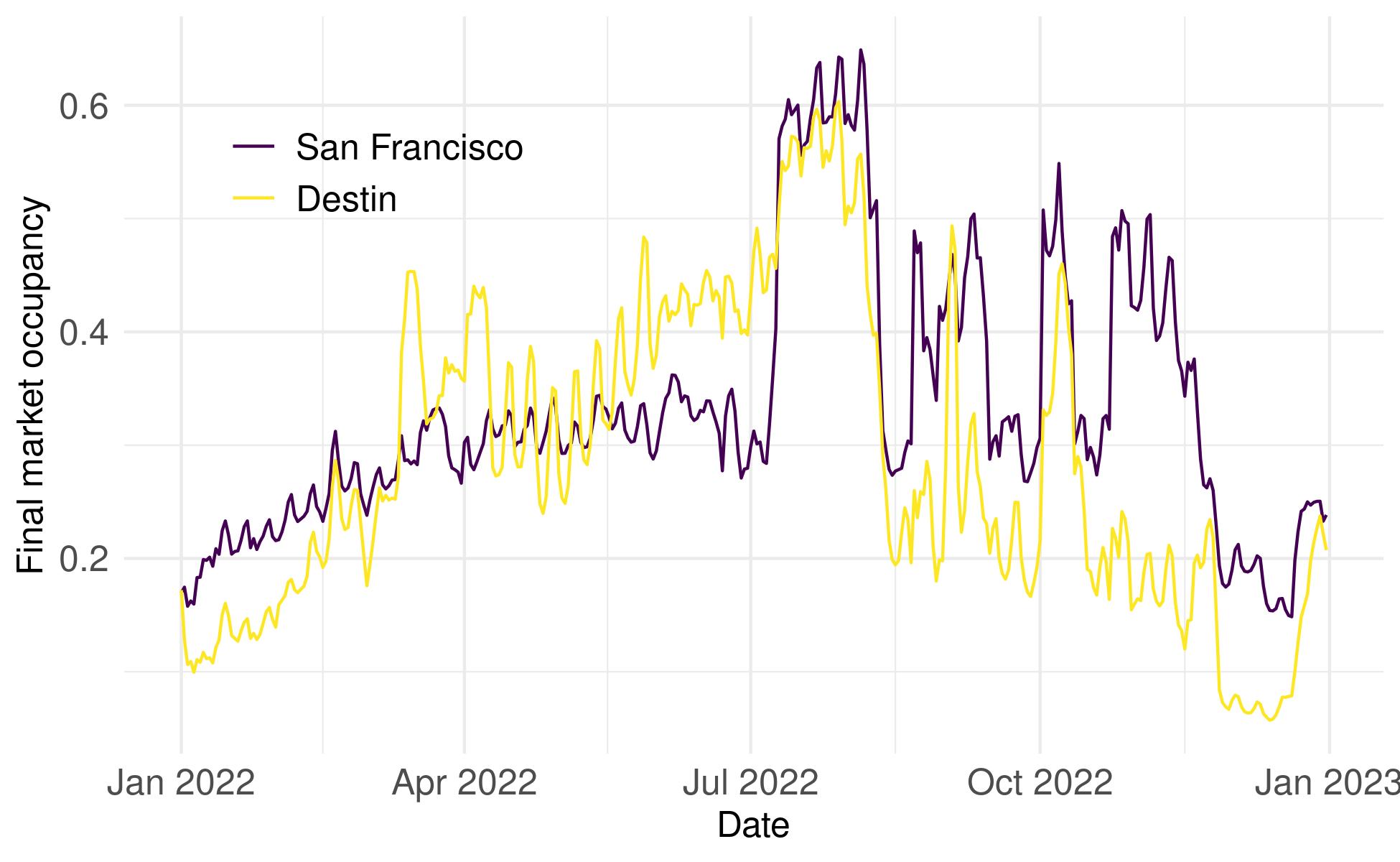} 
    \caption{Observed final market occupancy ($1-\mathbb{S}(t)$) in San Francisco, CA and Destin, FL for stay dates from January $1$st, $2022$ to December $31$st, $2022$. } 
    \label{fig:final_market_occ} 
\end{figure}

To further investigate the model performance, we compare the model performance at the market level for two markets with different dynamics; San Francisco, CA and Destin, FL. San Francisco possesses a typical city market behavior with weekend effects and large individual events, such as concerts, while Destin is a popular sunny vacation market. The differences in market behavior are illustrated by the observed final market occupancy in the test period shown in Figure \ref{fig:final_market_occ}. In San Francisco, the overall occupancy is high in summer and fall, with higher spikes during large events such as Outside Lands (early August). For Destin on the other hand, the occupancy is overall high during late spring, summer and early fall. Further spikes in occupancy are observed during holidays, such as during Easter. Due to the overall difference in the occupancy structure, how the prices should be adjusted dynamically will also be different. The pricing recommendations in typical city markets, such as San Francisco, are first adjusted to reflect day-of-the-week variations. In addition, major event such as the Outside Lands festival will substantially increase the prices. For Destin, the price recommendations are to a larger extent be adjusted to the season, with specific increase in price during different vacations and bank holidays.  Further examples comparing the stability of the forecasts are given in Appendix~\ref{app:stability}. 

\subsubsection{San Francisco}

\begin{figure}[H]
    \begin{subfigure}{0.49\linewidth}
        \includegraphics[page = 1, height=0.20\textheight, width=\linewidth]{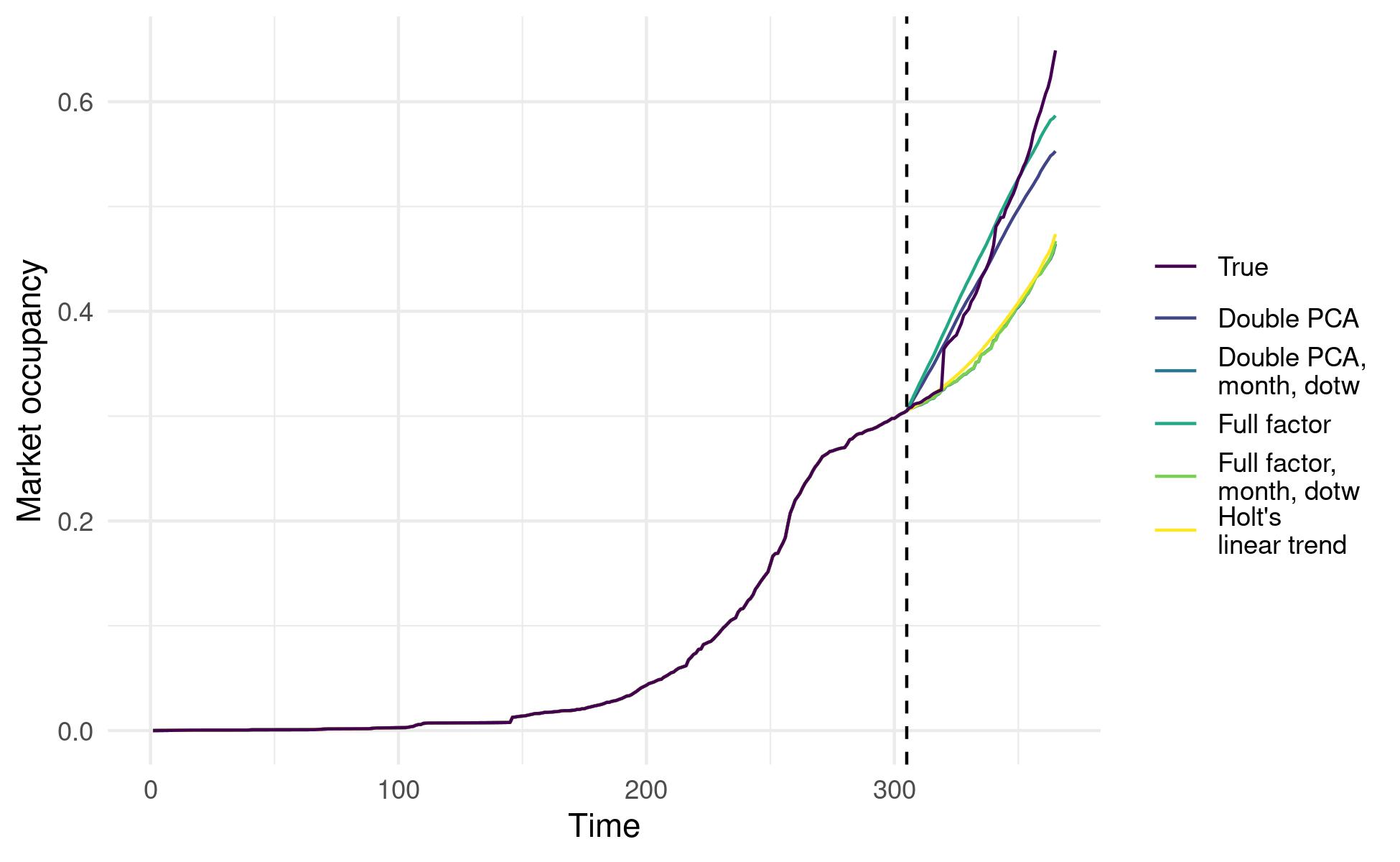}
        \caption{Forecast horizon of $60$ days.}
        \label{fig:pred_outside_lands_60}
    \end{subfigure}
    \hfill
    \begin{subfigure}{0.49\linewidth}
        \includegraphics[page = 1, height=0.20\textheight, width=\linewidth]{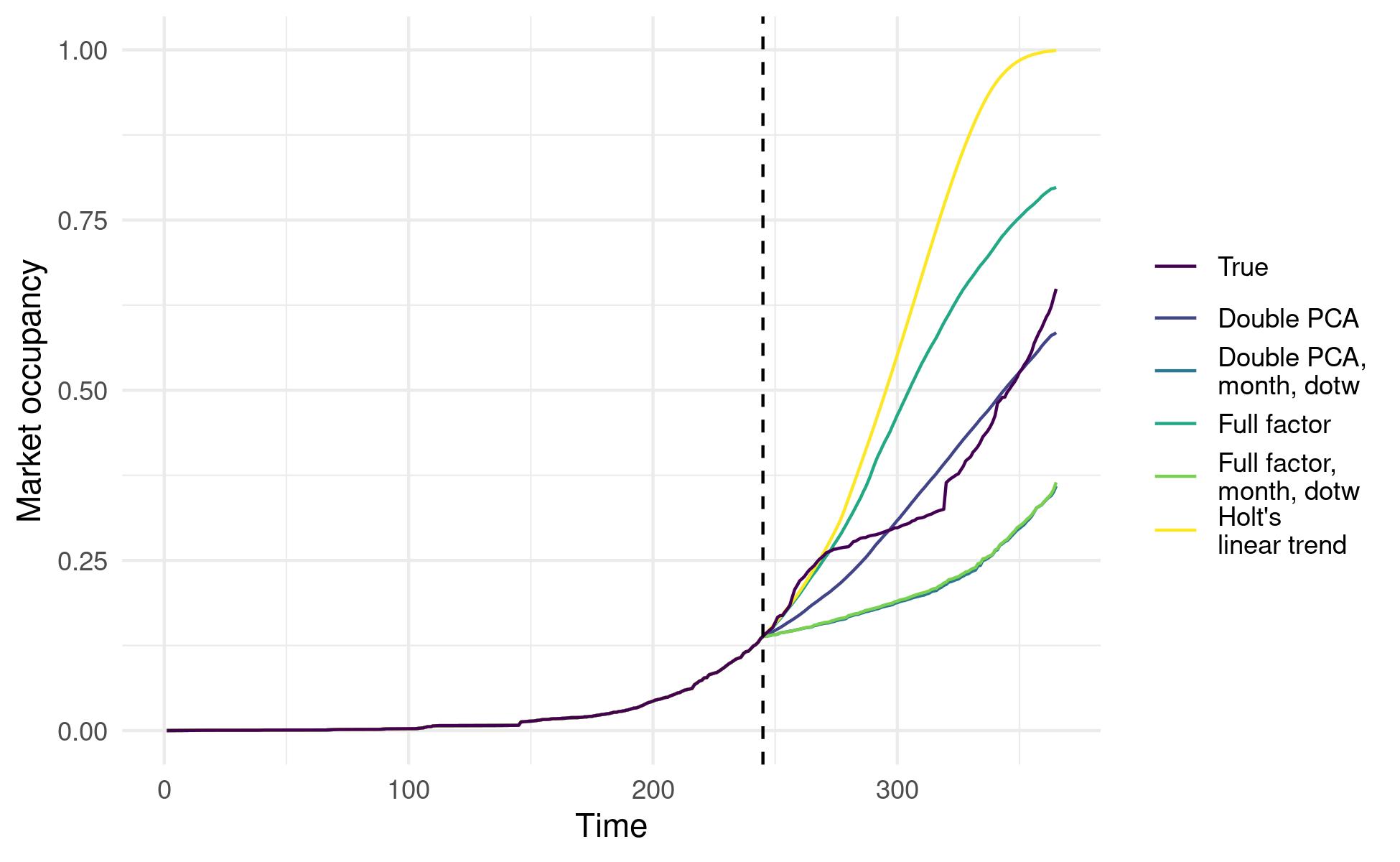}
        \caption{Forecast horizon of $120$ days.}
        \label{fig:pred_outside_lands_120}
    \end{subfigure}
    \caption{Predicted and observed market occupancy during Outside Lands (August $6$th $2022$) in San Francisco.}
    \label{fig:outside_lands_24}
\end{figure}

As an illustration of a forecast for San Francisco, Figure \ref{fig:outside_lands_24} shows the predicted market occupancy on August $6$th $2022$, the second day of the Outside Lands concert, for all five models, and a forecast horizon of $60$ days (Figure \ref{fig:pred_outside_lands_60}) and $120$ days (Figure \ref{fig:pred_outside_lands_120}). In this case, the predictions from the double PCA model are stable, and relatively accurate for both forecast horizons. While the full factor model yields less stable predictions, it shows the best performance $60$ days in advance. For both models, the performance deteriorates if the training data is stratified by month and day of the week and month. Finally, the predictions from Holt's linear trend model vary greatly in the two plots, yielding poor predictions in both cases.

Table \ref{tab:sf_perf_dotw_monthy_final30} compares the aggregated predictive performance of the five models over the test set of $2022$. We focus here on the last $30$ days of each forecast horizon, allowing for comparison across forecast horizons. The scores are overall somewhat higher than the aggregated scores over all markets listed in Table \ref{tab:overall_perf_last_30}, indicating that San Francisco is a challenging market to predict. One potential explanation is the challenge of predicting high-demand days due to the COVID-19 pandemic which is present in a substantial part of the training data. The pandemic altered typical event patterns because of cancellations and general market dynamics, thereby limiting the availability of historical booking trajectories representing high-demand market occupancy trajectories. 

\begin{center}
\captionof{table}{Mean IQD-values for San Francisco, aggregated over all stay dates in $2022$, for the final $30$ days of each forecast horizon. The best score for each forecast horizon is indicated in bold. If \ding{51}, day of the week and month effects are included in the model. }\label{tab:sf_perf_dotw_monthy_final30}
\begin{tabular}{l c P{1cm} P{1cm} P{1cm} P{1cm} P{1cm} P{1cm}} 
 \toprule
 & & \multicolumn{6}{c}{Forecast horizon} \\
\cline{3-8} 
Model & D/M & $30$ & $60$ & $90$ & $120$ & $150$ & $200$\\ 
 \toprule
 Full factor & \ding{55} & $0.060$ & $\bf 0.196$ & $0.397$ & $0.644$ & $0.956$ & $1.389$ \\
  Full factor & \ding{51} &$0.073$ & $0.242$ & $0.396$ & $0.580$ & $0.770$ & $\bf 0.889$\\
  Double PCA & \ding{55} & $0.062$ & $0.206$ & $\bf 0.365$ & $\bf 0.462$ & $\bf 0.751$ & $0.989$ \\
 Double PCA & \ding{51} &$0.075$ & $0.243$ & $0.395$ & $0.582$ & $0.791$ & $0.946$ \\
Holt & \ding{55} &$\bf 0.039$ & $0.259$ & $0.875$ & $1.475$ & $1.911$ & $2.777$\\
 \bottomrule
\end{tabular}
\end{center}

\subsubsection{Destin}
%\paragraph*{Point predictions for June $6$th $2022$}
\begin{figure}[H]
    \begin{subfigure}{0.49\linewidth}
        \includegraphics[page = 1, height= 0.20\textheight, width=\linewidth]{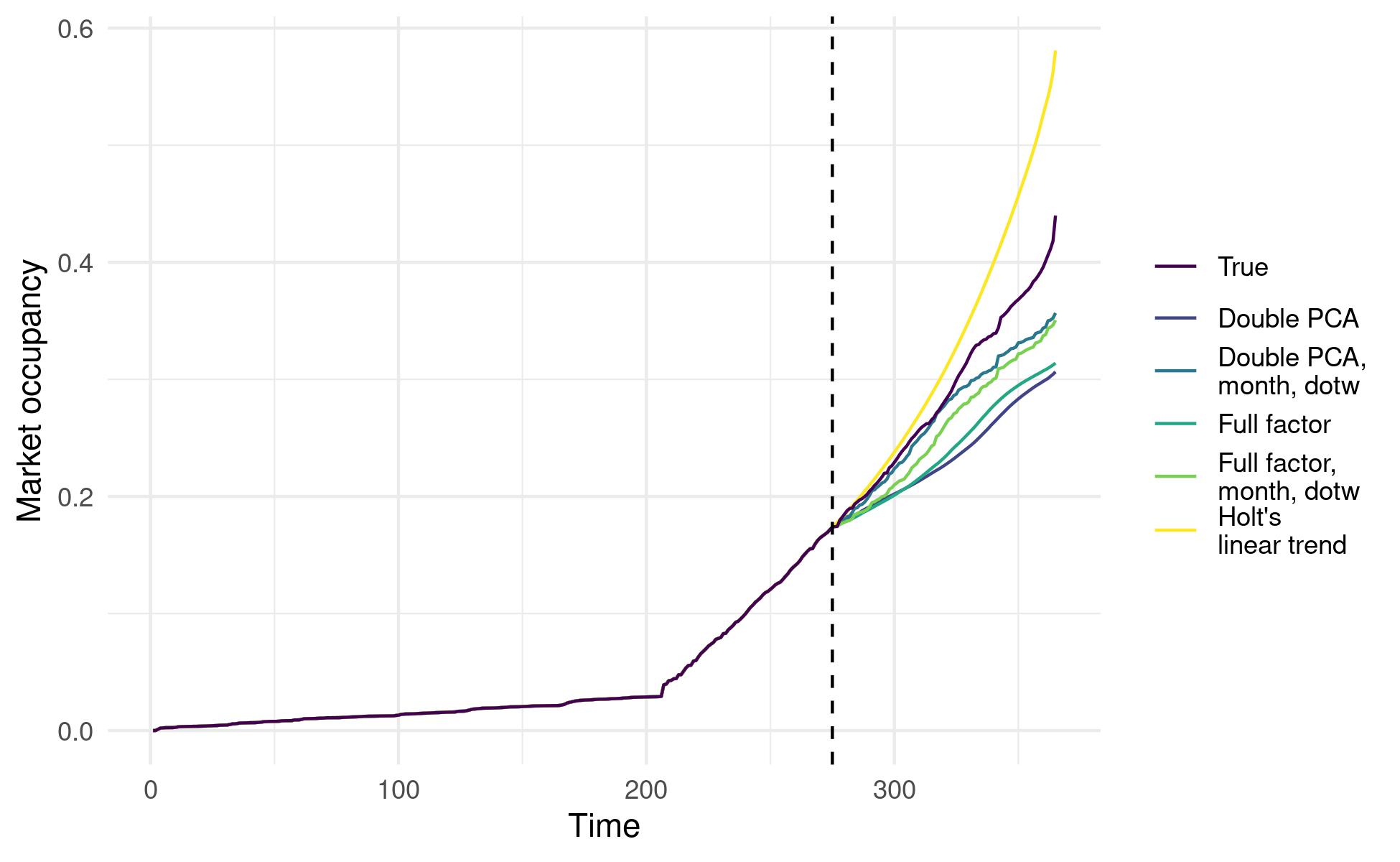}
        \caption{Forecast horizon of $90$ days.}
        \label{fig:jun_22_destin_a}
    \end{subfigure}
    \hfill
    \begin{subfigure}{0.49\linewidth}
        \includegraphics[page = 1, height= 0.20\textheight, width=\linewidth]{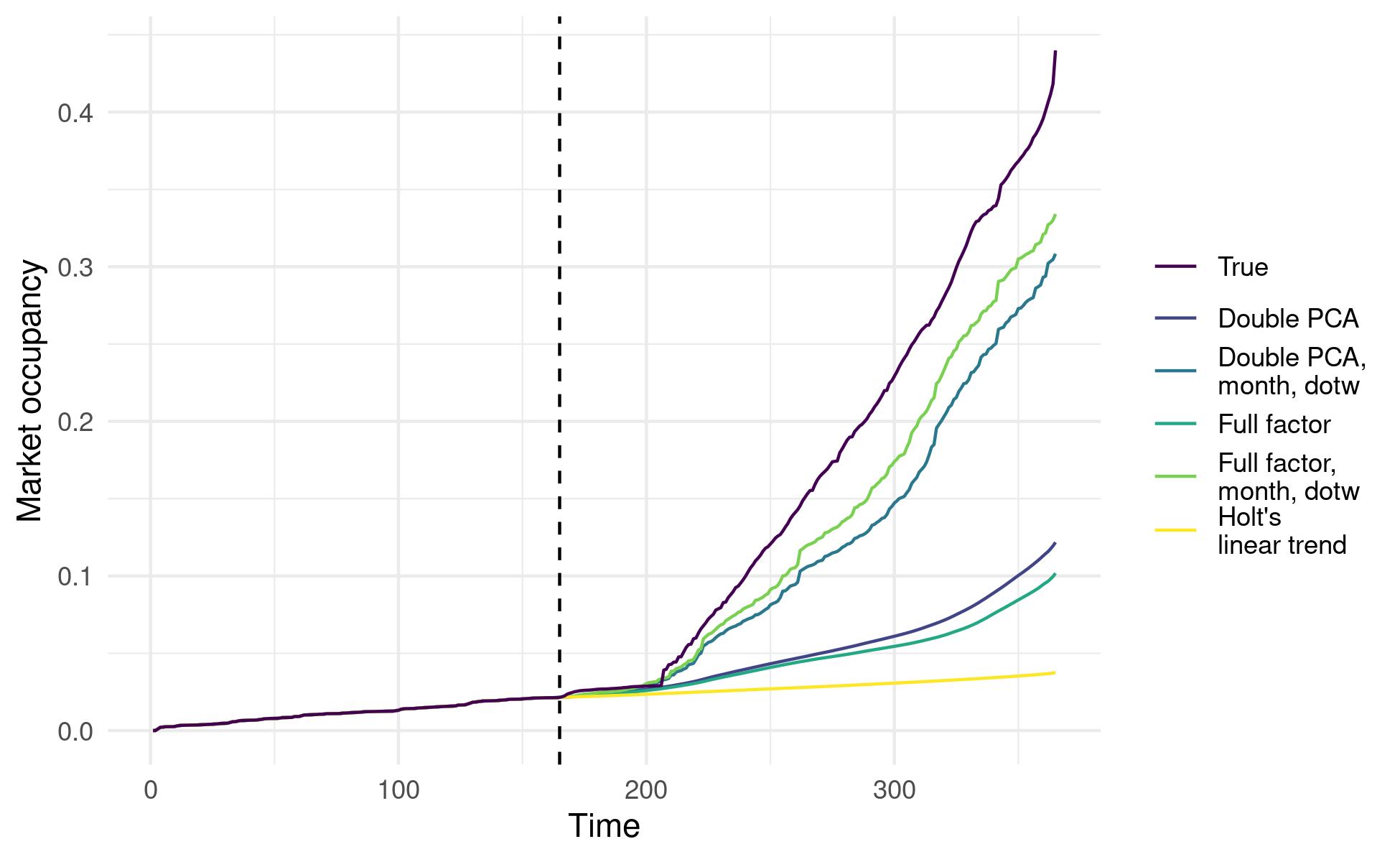}
        \caption{Forecast horizon of $200$ days.}
        \label{fig:jun_22_destin_b}
    \end{subfigure}
    \caption{Predicted and observed market occupancy in Destin on June $6$th $2022$.}
    \label{fig:jun_22_destin}
\end{figure}
We now turn our focus to Destin, FL. Figure \ref{fig:jun_22_destin} shows the predicted and observed market occupancy paths for June $6$th $2022$ using all five models. The true market occupancy line is low, almost flat, in the beginning, indicating a potential low-demand date. However, at around $150$ days before the stay date, the market occupancy curve increases fast. If a host, for example, uses the forecasts from Holt's linear trend model $200$ days before the stay date, a low price would be recommended for this date. However, since the PCA-based models take the historical data into account, early predictions from the models with day of the week and month effects can quite accurately forecast the final path. The host can therefore set the price higher according to the actual demand patterns, ending up with a higher revenue.  

We continue by examining the performance of the models for all the stay dates in the test set in Destin. The resulting mean IQD-values are shown in Table \ref{tab:destin_perf_dotw_monthy_final30} for the final $30$ days of the forecast horizon. Including day of the week and month clearly results in a lower mean IQD-value for both the full factor analysis model and the double PCA model when the forecast horizon is longer than $60$ days. This indicates a different market behavior compared to San Francisco. The performance of the models for San Francisco, CA. Table \ref{tab:sf_perf_dotw_monthy_final30}, showed limited benefits of including day of the week and month effects, especially for double PCA and for shorter forecast horizons. However, in Destin, inclusion of day of the week and month effects results in considerably better predictions even for shorter forecast horizons. When a host sets the price for a unit, it should be dynamically adjusted to capture the temporal demand patterns associated with peak and off-peak seasons. This may more important for sunny vacation markets such as Destin, compared to a larger city such as San Francisco.

\begin{center}
\captionof{table}{Mean IQD-values for Destin, aggregated over all stay dates in $2022$, for the final $30$ days of each forecast horizon. The best score for each forecast horizon is indicated in bold. If \ding{51}, day of the week and month effects are included in the model.}\label{tab:destin_perf_dotw_monthy_final30}
\begin{tabular}{l c P{1cm} P{1cm} P{1cm} P{1cm} P{1cm} P{1cm}} 
 \toprule
 & & \multicolumn{6}{c}{Forecast horizon} \\
\cline{3-8} 
Model & D/M & $30$ & $60$ & $90$ & $120$ & $150$ & $200$\\ 
 \toprule
 Full factor & \ding{55} & $0.074$ & $0.213$ & $0.379$ & $0.574$ & $0.646$ & $1.311$ \\
  Full factor & \ding{51} & $0.074$ & $0.214$ & $0.314$ & $0.367$ & $\bf 0.389$ & $\bf 0.465$ \\
  Double PCA & \ding{55} & $0.072$ & $0.219$ & $0.384$ & $0.540$ & $0.671$ & $0.921$ \\
 Double PCA & \ding{51} & $0.076$ & $0.211$ & $\bf 0.308$ & $\bf 0.364$ & $0.408$ & $0.527$\\
Holt & \ding{55} & $\bf 0.023$ &  $\bf 0.123$ & $0.316$ & $0.670$ & $1.329$ & $2.935$\\
 \bottomrule
\end{tabular}
\end{center}

\section{Conclusions}\label{sec:Conclusion}
This article develops two prediction models for partially observed survival curves that combine fully observed historical curves with the currently observed partial curve. Both models are built on principal component analysis: a full factor analysis model and a double PCA model. We also show how to incorporate calendar effects through day‑of‑week and month indicators.

The models are motivated by a recently released dataset on the short-term rental market. The data consists of time series of market-level occupancy curves for $500$ markets from $2017$ through $2022$, where each stay date is represented by a $366$-point survival path over the booking horizon. This yields a high-dimensional time series of strongly correlated survival curves, for which new methods are needed to capture the dependence structure.

We evaluate performance using the integrated quadratic distance (IQD) of \citet{thorarinsdottir2013using}. For each forecast horizon, we compute the mean IQD for each market over all stay dates in $2022$. The results show that our models produce accurate and stable forecasts across horizons. We also compare the PCA-based models to Holt's linear trend model and find that the PCA-based models generally outperform Holt's model for forecast horizons longer than about $60$ days. 

We further examine how performance varies with the forecast horizon across markets. For example, when the forecast horizon increases from $60$ to $120$ days, Holt's linear trend model becomes universally inferior. We illustrate the behavior of the models in detail for specific markets, such as San Francisco and Destin, FL, where we present and discuss both point predictions and IQD performance. In these examples, day of the week and month effects are less important in San Francisco, but substantially improve performance in Destin.  

Although the models are developed for survival curves, they are not restricted to this context. We work in real space by transforming survival curves via a time-varying Type-II Lehmann model, yielding a general framework for curve-based data with many potential applications in functional data analysis.  

Obvious related business applications include airline bookings, while non-business applications include hospital bed occupancy. In such settings, some adaptation of the framework will likely be necessary to respect the specifics of the data. There are many possibilities for extension; for example, one can enhance the framework to share information across markets or time points through the covariance structure. This would allow borrowing strength across related series--for instance, using experience from other markets or previous years to improve forecasts around special events such as Outside Lands. 

\section{Acknowledgments}\label{Acknowledgments}
The authors acknowledge the support of the Research Council of Norway, MEA through grant 342613 "Predicting in High Dimensions" and TLT through the Centre of Excellence "Integreat – The Norwegian Centre for Knowledge-driven Machine Learning", project number 332645.  

\section{Declaration of interest}\label{disclosure-statement}
The authors do not have any known conflicts of interest. 

\section{Data availability statement}\label{data-availability-statement}
The data with a thorough description has been released in \citet{datapaper}. 

\bibliography{literature.bib}

%\printbibliography %Prints bibliography

\newpage
\phantomsection\label{supplementary-material}
\bigskip

\begin{center}

{\large\bf SUPPLEMENTARY MATERIAL}

\end{center}

\begin{description}
\item[Title: Background models]
Some mathematical transformations for the background models.
\item[Title: Dynamically estimating $\mathbb{S}_0$]
How $\mathbb{S}_0$ is estimated dynamically.
\item[Title: Detailed analysis for dimension selection]
A detailed analysis of the decision of the number of principal components. 
\item[Title: Stability of the predictions] 
Individual IQD-values for forecast horizon $30$ versus $60$, which shows the stability of the predictions.
\item[R-code: Code for all results] To be inserted after blind review.
\end{description}

\clearpage
\begin{appendices}

\section{Background models} \label{app: background}
The baseline proportion $\mathbb{S}_0(t)$ is defined as
\begin{align*}
    \mathbb{S}_0(t) = \prod_{r\in\mc{T}: r\leq t}\xi_{0r}.
    \label{eq:surv_0}
\end{align*}
In addition from Equation \eqref{eq:surv_1}, we have
\begin{align*}
    \mathbb{S}_i(t) & = \prod_{r\in\mc{T}: r\leq t}\xi_{ir}\\
    & = \prod_{r\in\mc{T}: r\leq t}(\xi_{0r})^{exp(\gamma_{ir})} \\
    & = \xi_{0t}^{exp(\gamma_{it})} \mathbb{S}_i(t-1).
\end{align*}
Rearranging this, we get
\begin{align*}
    \frac{\mathbb{S}_i(t)}{\mathbb{S}_i(t-1)} & = \xi_{0t}^{exp(\gamma_{it})} \\
    & = \Big(\frac{\mathbb{S}_0(t)}{\mathbb{S}_0(t-1)}\Big)^{exp(\gamma_{it})},
\end{align*}
and we end up with Equation \eqref{eq:gamma_t}

In some situations, we have that $\mathbb{S}_i(t) = \mathbb{S}_i(t-1)$. We let the value of $\widehat{\gamma}_{it}$ be constant for these values. We get
\begin{align*}
     \mathbb{S}_i(t) & = \prod_{r\in\mc{T}: r\leq t}\xi_{ir} \\
     & = \Big( \prod_{r\in\mc{T}: r\leq v-1}\xi_{0r}^{exp(\gamma_{ir})} \Big)\Big( \prod_{r\in\mc{T}: v \leq r\leq v+h}\xi_{0r}^{exp(\gamma_{ir})} \Big)\Big( \prod_{r\in\mc{T}: v+h+1 \leq r\leq t}\xi_{0r}^{exp(\gamma_{ir})} \Big).
\end{align*}
If $\gamma_{ir}$ is constant when $r$ is between $v$ and $v+h$. then $\gamma_{ir} = \widehat{\gamma}_i$ and we get that 
\begin{align*}
\gamma_{i} = \log\left(\frac{\log(\mathbb{S}_{i}(v+h) / \mathbb{S}_{i}(v))}{\log(\mathbb{S}_0(v+h) / \mathbb{S}_0(v)}\right).
\end{align*}

For example if $\widehat{\mathbb{S}}_{i}(49) \neq \widehat{\mathbb{S}}_{i}(50) = \widehat{\mathbb{S}}_{i}(51) = \widehat{\mathbb{S}}_{i}(52) \neq \widehat{\mathbb{S}}_{i}(53)$, we get that 
\begin{equation*}
    \tilde{\gamma}_{i} = \log \Bigg(\frac{\log \Big( \widehat{\mathbb{S}}_{i}(53)/\widehat{\mathbb{S}}_{i}(50) \Big)}{\log \Big(\widehat{\mathbb{S}}_0(53)/\widehat{\mathbb{S}}_{0}(50)\Big)} \Bigg)
\end{equation*}
and this value applies for $\mathbb{S}_{i}(51)$ and $\mathbb{S}_{i}(52)$. 

\section{Dynamically estimating $\mathbb{S}_0$}
\label{sec:S_0}
When performing dynamical forecasting, $\mathbb{S}_0$ can not be static, since the training and test data changes. Therefore, $\mathbb{S}_0(t)$ is estimated from the available training data. Following \citet{datapaper_blind}, the estimated survival curve can be written as
\begin{align*}
    \mathbb{S}_0(t) & = \prod_{r\in\mc{T}: r\leq t} \Big( 1 - \frac{d_{0r}}{n_{0r}} \Big) \\
    & = \prod_{r\in\mc{T}: r\leq t}\xi_{0r},
\end{align*}
where $d_{0r}$ is the estimated number of bookings that happened at time $r$ for all the stay dates in the training data, and $n_{0r}$ is the estimated number of listings that have survived up to time $r$ for all the stay dates in the training data. If the training data is for $k$ number of stay days, then
\begin{align*}
    d_{0r} & = \sum_{i = 1}^{k} d_{ir}, \\
    n_{0r} & = \sum_{i = 1}^{k} n_{ir}.
\end{align*}

\section{Detailed analysis for dimension selection}
\label{app:dimension selection}
\begin{figure}[H]
\centering
        \includegraphics[page = 1, height=0.25\textheight, width=0.9\linewidth]{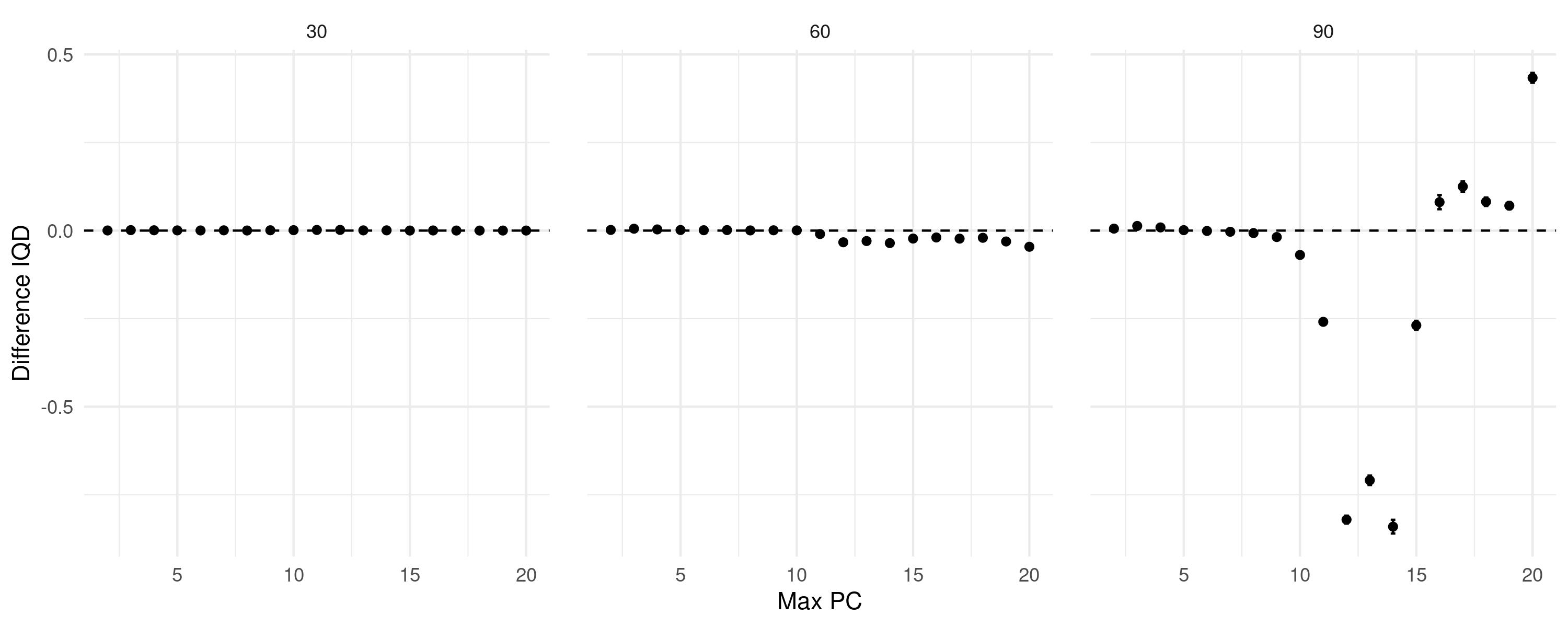}
    \caption{Increments in aggregated IQD-values for dimensions $n = 2, \dots, 20$ for forecast horizons of $30, 60$ and $90$ days for the full factor model. The results are based on a $10$-fold cross validation on data from $2018$ over all $500$ markets.}
    \label{fig:iqd_diff_factor_zoom}
\end{figure}
Here, we provide a detailed explanation of the dimension-selection results in Table \ref{tab:n_pc}. Our focus is on the setting where a single model is estimated over all days in the test set. Even though this method should clearly give an indication of the number of principal components, in practice it can fail to do so. In such cases, we select the number of principal components to correspond to the nearest forecast horizon where the results exhibit a clear pattern. 

Figure \ref{fig:iqd_diff_factor_zoom} shows the increments in aggregated IQD-values, $IQD_{diff}^{n-1} = IQD_{n-1} - IQD_n$, for forecast horizons of $30$, $60$ and $90$ days. Since the increments cluster near $0$, the number of principal components for forecast horizon $30$ and $60$ seems more or less immaterial. For a forecast horizon of $90$ days however, we observe a negative effect on model performance when including more than $10$ principal components. In addition, the point are marginally above the zero line until we reach $5$ principal components. For this forecast horizon, we choose to include $4$ principal components. Since the number of principal components for forecast horizon $30$ and $60$ seems more or less immaterial, we set these numbers equal to when the forecast horizon is $90$ days. 

\begin{figure}[H]
\centering
        \includegraphics[page = 1, height=0.25\textheight, width=0.9\linewidth]{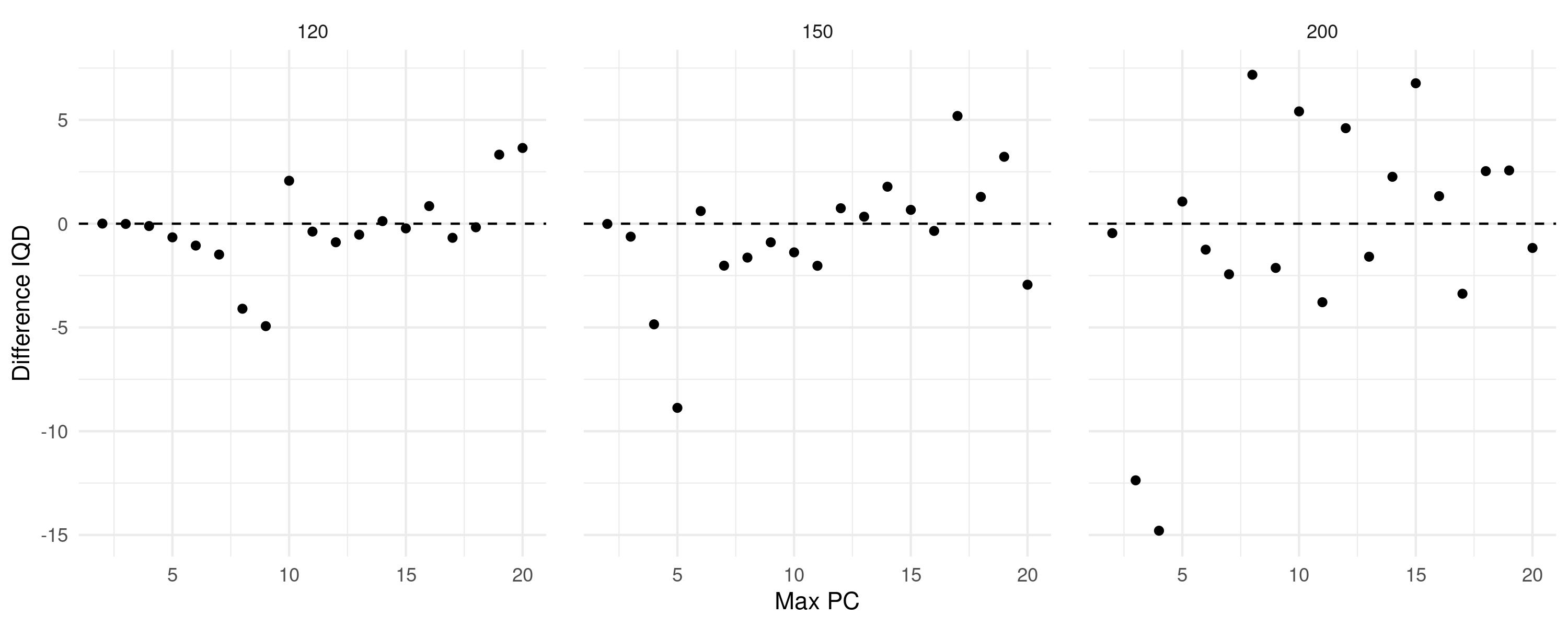}
    \caption{Increments in aggregated IQD-values for dimensions $n = 2, \dots, 20$, forecast horizons of $120, 150 $ and $200$ days for the full factor model. The results are based on a $10$-fold cross validation on data from $2018$ over all $500$ markets.}
    \label{fig:iqd_diff_factor}
\end{figure}

The corresponding results for forecast horizons of $120$, $150$ and $200$ days are shown in Figure \ref{fig:iqd_diff_factor}. For these forecast horizons, the results are more sensitive to the model dimension compared to lower forecast horizons. When the forecast dimension is $120$ days, the first time the line is crossed is when the number of principal components is $4$. This means it is not beneficial to include the fourth principal component, and we therefore include $3$ principal components. The same argument applies when the forecast dimension is $150$, and it results in $2$ principal components. For a forecast horizon of $200$ days, the results provide no clear indication of how many principal components should be included since the zero line is already crossed at $2$. For this case, we therefore look at when the number of principal components are above the zero line again. That happens when at $5$. Again, the line is crossed at $6$ principal components, and we therefore choose to include $5$ principal components. 

\begin{figure}[H]
\centering
        \includegraphics[page = 1, height=0.5\textheight, width=0.9\linewidth]{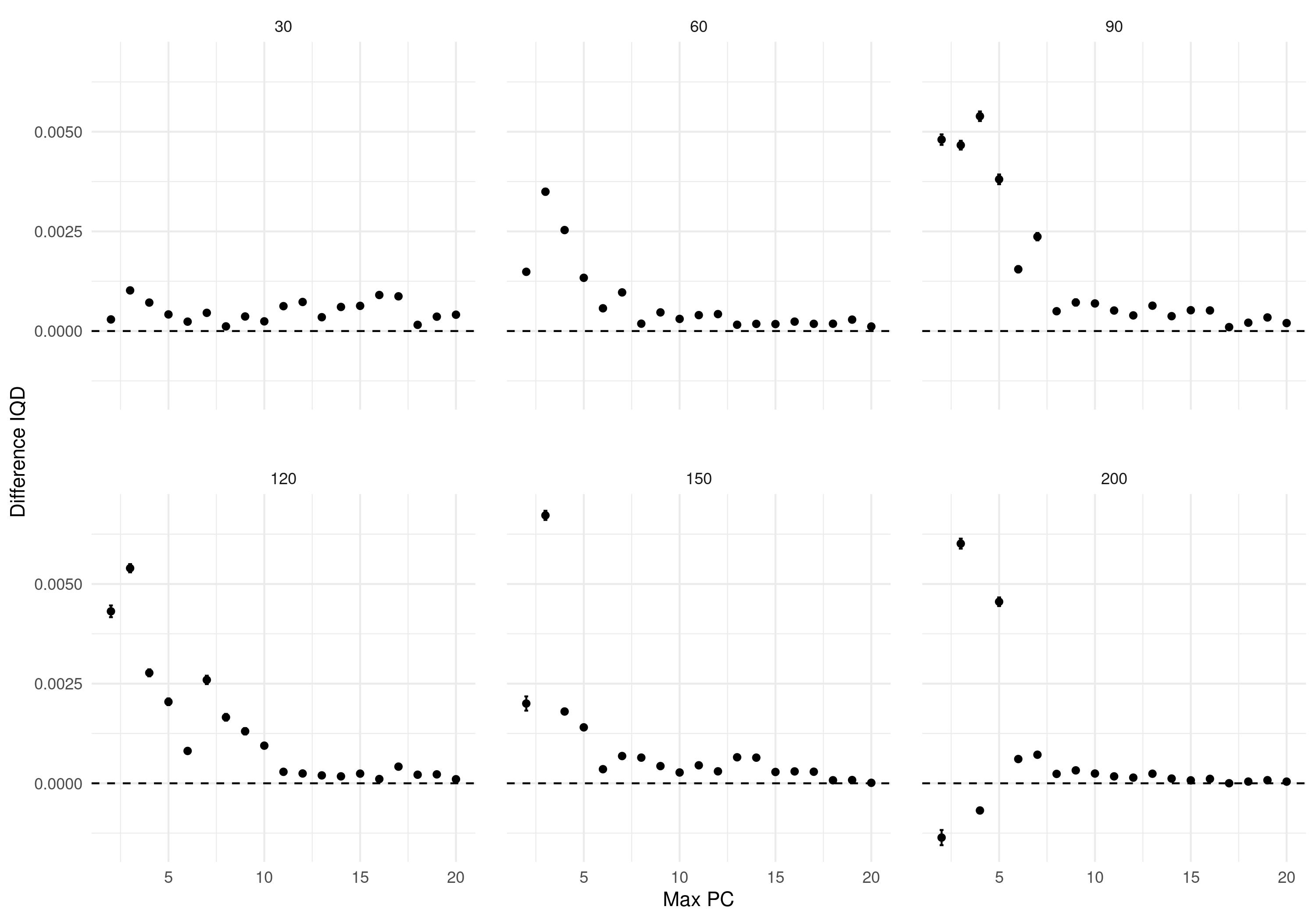}
    \caption{Increments in aggregated IQD-values for dimensions $n = 2, \dots, 20$, and forecast horizons of $30, 60, 90, 120, 150 $ and $200$ days for the double PCA model assuming $D_1 = D_2$. The results are based on a $10$-fold cross validation on data from $2018$ over all $500$ markets.}
    \label{fig:iqd_diff_double_pca}
\end{figure}

For the double PCA model, we initially assume equal number of principal components in the training and test period of each curve, that is, $D_1 = D_2$ in Equations \eqref{eq:gamma_A} and \eqref{eq:gamma_B}. As shown in Figure \ref{fig:iqd_diff_double_pca}, the model is highly robust to the number of principal components and only minimal differences in performance are observed across all forecast horizons. Furthermore, we see that, as opposed to the full factor model, the double PCA model seems robust against overfitting in that the increments stabilize at around zero as the dimensionality grows. For each forecast horizon, we select the dimensionality as listed in Table \ref{tab:n_pc} based on where the increments become essentially identical. 

We further investigated the impact of relaxing the assumption that $D_1=D_2$. This yielded only limited gains in predictive performance (results not shown). We therefore perform the main analysis with $D_1=D_2$.

\section{Stability of the predictions} \label{app:stability}
\begin{figure}[H]
\centering
        \includegraphics[page = 1, height=0.3\textheight, width=0.7\linewidth]{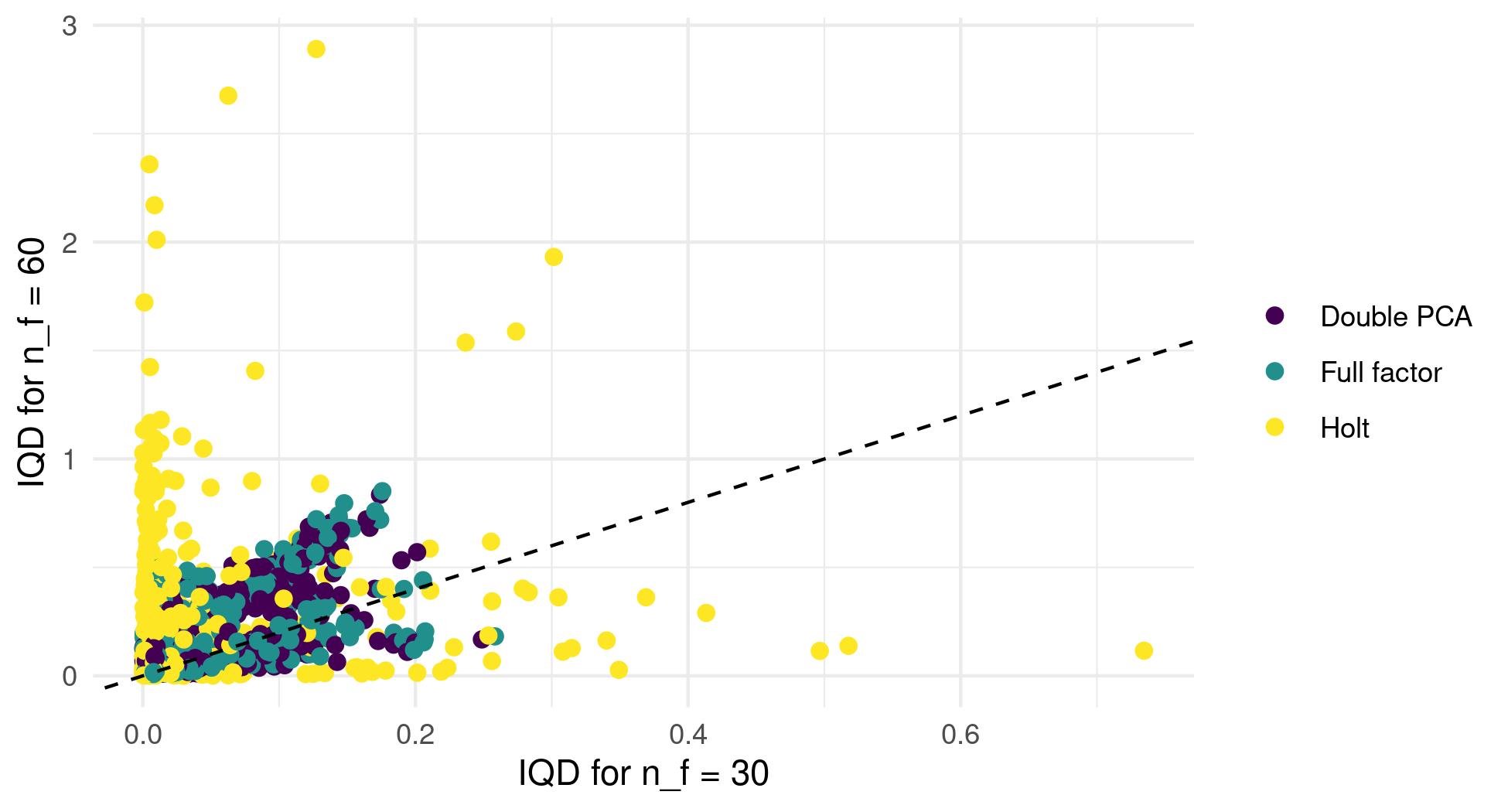}
    \caption{Individual IQD for January $1$st $2022$ to December $31$st $2022$ for forecast horizon $30$ versus $60$ in San Francisco. The dashed line has a slope of $2$.}
    \label{fig:iqd_30_60_market1}
\end{figure}
One of the advantages of double PCA and full factor analysis is the stability of the forecasts over the different forecast horizons. This is illustrated in Figure \ref{fig:iqd_30_60_market1}. The x-axis is IQD for a forecast horizon of $30$ days, while the y-axis is IQD for a forecast horizon of $60$ days. Each point represent the IQD-value for a stay date between January $1$st $2022$ to December $31$st $2022$. For full factor analysis and double PCA, the IQD-values are in most cases improving when the forecast horizon is lower. However, for Holt's linear trend model, the IQD for forecast horizon of $30$ days can quite often be higher than for a forecast horizon of $60$ days. This is because the weight of the recent observations are much higher compared to the other two methods.

\begin{figure}[H]
    \begin{subfigure}{0.5\linewidth}
        \includegraphics[page = 1, height= 0.25\textheight, width=\linewidth]{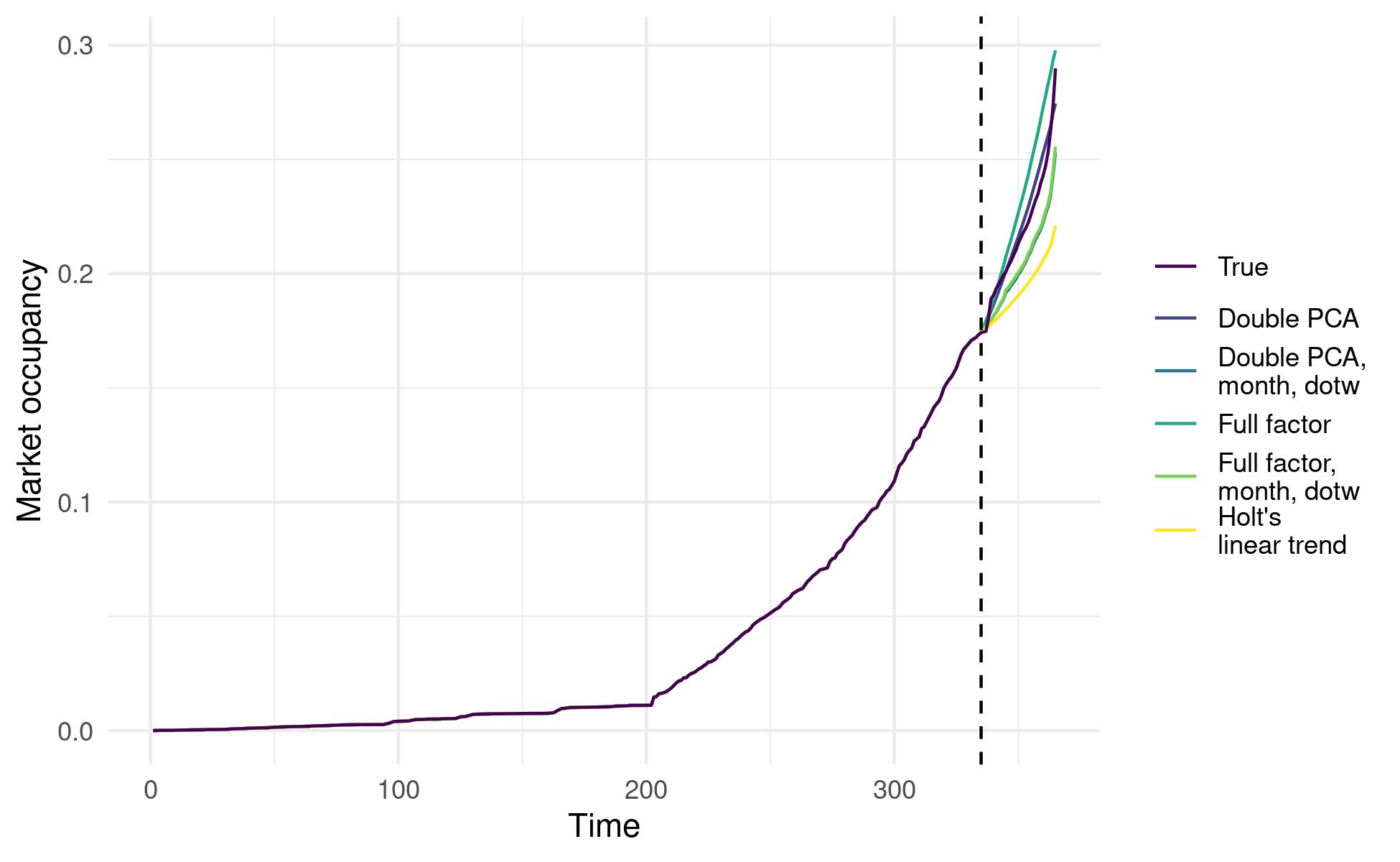}
        \caption{Forecast horizon of $30$ days.}
        \label{fig:aug_24_tahoe_a}
    \end{subfigure}
    \hfill
    \begin{subfigure}{0.5\linewidth}
        \includegraphics[page = 2, height= 0.25\textheight, width=\linewidth]{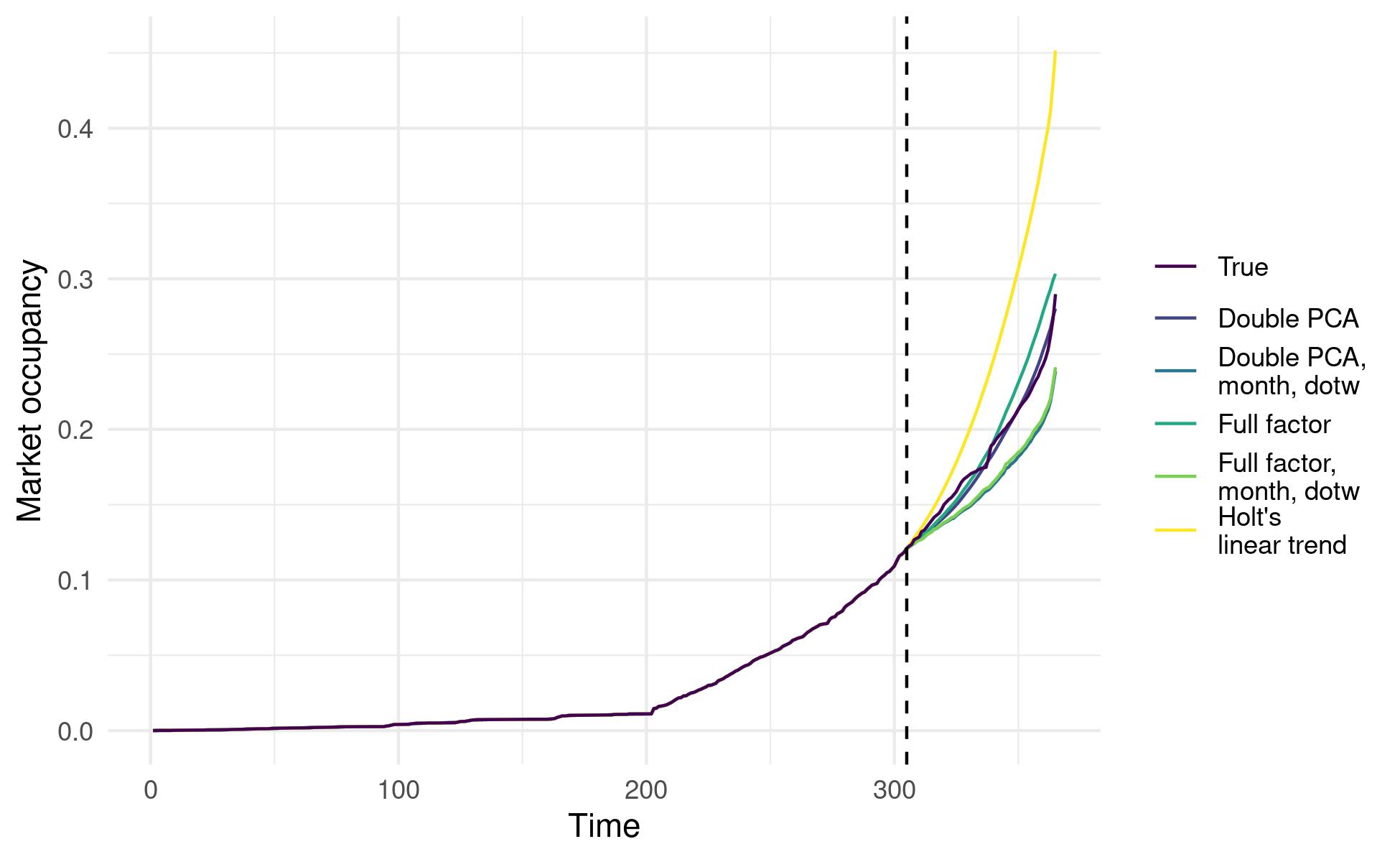}
        \caption{Forecast horizon of $60$ days.}
        \label{fig:aug_24_tahoe_b}
    \end{subfigure}
    \caption{Forecasting June $10$th $2022$ in Lake Tahoe using double PCA (with and without day of the week and month), full factor analysis (with and without day of the week and month) and Holt's linear method.}
    \label{fig:aug_24_tahoe}
\end{figure}

A further comparison of the stability of the different forecasts may be illustrated on the following example from Lake Tahoe in California and Nevada. In Figure \ref{fig:aug_24_tahoe}, we predict the market occupancy curves for June $10$th $2022$ using the five forecasting models $60$ and $30$ days before the stay date. All of the PCA-based models in this example are stable and with high performance. However, Holt's linear trend model has weak predictive performance and, more notably, highly unstable predictions. Figure \ref{fig:aug_24_tahoe} thus illustrates the sensitivity of Holt's linear trend model to recent observations. At $60$ days before the stay date, Holt's linear trend model is elevated, tracking well-above the truth. However, the prediction changes dramatically $30$ days before the stay date. With more observations, the final market occupancy decreases by over $20$ percentage points and the path is well below the true market occupancy path. The example illustrates that this models may be too sensitive to recent observations. 

\end{appendices}

\end{document}